\pdfoutput=1
\documentclass{article}

\usepackage{PRIMEarxiv}

\usepackage[utf8]{inputenc}
\usepackage[T1]{fontenc}
\usepackage[numbers,sort&compress]{natbib}
\usepackage{hyperref}
\usepackage{url}
\usepackage{booktabs}
\usepackage{amsfonts}
\usepackage{nicefrac}
\usepackage{microtype}
\usepackage{fancyhdr}
\usepackage{graphicx}

\usepackage{amsmath}
\usepackage{amssymb}
\usepackage{xcolor}
\usepackage{pgfplots}
\pgfplotsset{compat=1.18}
\usepackage{array}
\usepackage{tabularx}
\newcolumntype{L}[1]{>{\raggedright\arraybackslash}p{#1}}

\def\keywordname{\textbf{\emph{Keywords}}}

\newcommand{\PAI}{P_{AI}}

\definecolor{seriesA}{HTML}{2A78D6}
\definecolor{seriesB}{HTML}{EB6834}
\definecolor{inkgray}{HTML}{52514E}
\definecolor{caseFunded}{HTML}{1BAF7A}
\definecolor{caseAvoid}{HTML}{EDA100}

\newsavebox{\takeawaysavebox}
\newenvironment{takeaway}[1]
 {\par\smallskip\noindent\setlength{\fboxsep}{6pt}%
  \begin{lrbox}{\takeawaysavebox}%
  \begin{minipage}{\dimexpr\linewidth-2\fboxsep-2\fboxrule\relax}%
  \textbf{#1.}\ \ignorespaces}
 {\end{minipage}\end{lrbox}%
  \fcolorbox{seriesA}{seriesA!6}{\usebox{\takeawaysavebox}}\par\smallskip}

\newsavebox{\examplesavebox}

\title{Today's Work, Tomorrow's Workforce:\texorpdfstring{\\}{ } Incentivizing Learning in AI-Assisted Work}

\newcommand{\authorcell}[4]{\begin{tabular}[t]{@{}c@{}}\textbf{#1}\\ #2\\ #3\\ \texttt{#4}\end{tabular}}
\author{\mdseries
\begin{tabular}{@{}>{\centering\arraybackslash}p{0.45\textwidth}>{\centering\arraybackslash}p{0.45\textwidth}@{}}
\authorcell{Moritz Diener\thanks{Equal contribution.}}{Karlsruhe Institute of Technology (KIT)}{Karlsruhe, Germany}{moritz.diener@kit.edu} &
\authorcell{Daniel Hendriks\footnotemark[1]}{Karlsruhe Institute of Technology (KIT)}{Karlsruhe, Germany}{daniel.hendriks@kit.edu} \\ \noalign{\vspace{2.2ex}}
\authorcell{Philipp Spitzer}{Karlsruhe Institute of Technology (KIT)}{Karlsruhe, Germany}{philipp.spitzer@kit.edu} &
\authorcell{Gerhard Satzger}{Karlsruhe Institute of Technology (KIT)}{Karlsruhe, Germany}{gerhard.satzger@kit.edu} \\ \noalign{\vspace{2.2ex}}
\authorcell{Joshua Holstein}{Karlsruhe Institute of Technology (KIT)}{Karlsruhe, Germany}{joshua.holstein@kit.edu} &
\authorcell{Wei Sun}{IBM Research}{USA}{sunw@us.ibm.com}
\end{tabular}}

\begin{document}
\maketitle

\begin{abstract}
At work, people develop skills through experience: novices acquire skills by completing tasks, while experts retain theirs through continued practice. AI assistance changes this process by offering workers a choice between solving tasks independently and accepting the AI's answers.  While relying on AI saves immediate effort, it can also displace the practice needed to acquire and retain skills, including those required to oversee the AI. We extend an existing accept-or-solve model to a sequence of tasks in which workers' skills evolve through independent practice and reliance on AI. We compare myopic workers, who maximize immediate utility, with strategic workers, who also value their future skills. We show that novices can become trapped at low skill levels, while experts may lose skill, even as AI-assisted outcome accuracy remains high. Strategic workers, by contrast, may choose to solve independently even when AI is more accurate, recognizing the future value of acquiring and retaining their skills. We then study incentives that reward correct independent work and characterize the payments needed to encourage skill acquisition and retention. We show that independent work can become self-sustaining once workers acquire sufficient skill, eliminating the need for continued incentives. Counterintuitively, supporting a higher skill target can sometimes reduce long-run incentive costs, either by enabling self-sustaining independent work or by lowering the payments needed to maintain proficiency. Our findings thereby shift the focus from whether workers should rely on AI on individual tasks to how organizations can sustain human expertise over time, and which skills are worth investing in.
\end{abstract}


\section{Introduction}\label{sec:intro}

\textit{Astronauts on a six-month mission aboard the International Space Station need little physical strength for their work, but without gravity and physical effort, muscles and bones deteriorate. Since no work in space requires this physical strength, this loss often goes unnoticed until astronauts return to Earth. Once this loss becomes evident, space agencies began scheduling about two and a half hours of exercise into every mission day. This exercise retains the capacity that astronauts need once gravity returns. What gravity and physical effort are for an astronaut's body is cognitive effort for a worker's skills.}

Traditionally, work has trained the people who perform it: novices acquire skill by solving tasks, and experts retain their skill through regular use \cite{billett2001, lave1991, ericsson1993}. AI loosens this link as workers can now accept an AI's answer instead of solving a task themselves \cite{chalkidis2026}, often finishing faster and with higher quality \cite{dellacqua2026}. The effort that AI saves, however, is the very effort through which workers learn. For example, novices who relied on AI while learning a new programming library acquired less skill \citep{shen2026}, and endoscopists who had worked with AI detected fewer adenomas once the AI was withdrawn \citep{budzyn2025}. Work performed with AI assistance therefore risks workers not acquiring or retaining skill.

Even as AI capabilities grow, acquiring and retaining skill remains important because skills build on each other and are therefore needed to acquire other skills \cite{gagne1962}. For example, programmers may use AI for routine coding yet still need coding skill to acquire additional skills that enable them to debug failures and make architectural judgments. This matters most early in a career, when novices perform routine tasks to build the foundation for becoming an expert \cite{brynjolfsson2026canaries, hosseini2026}. However, the importance of skills also extends to a longer horizon, where it is needed to oversee the AI itself, i.e., to detect its errors and take over when it fails \cite{bainbridge1983,jain2026}. For high-risk systems, oversight is even legally mandatory through the EU AI Act  \cite{euaiact2024}.

Research on human--AI decision-making has evaluated workers' choice to solve alone or use an AI by \emph{appropriate reliance}: a worker should accept the AI's answer when it is more likely to be correct and solve the task independently otherwise \cite{leesee2004, schemmer2023}. This accept-or-solve choice is represented in formal models by maximizing the utility of a worker on a single task \cite{bansal2021, holstein2026}. In particular, \citet{holstein2026} show that independent effort can make accepting the AI's answer rational even when the worker is more skilled and therefore more likely to answer correctly. With skill held fixed, however, these models miss two dynamics: A novice acquires skill by solving tasks that the AI would answer more accurately, which appropriate reliance scores as underreliance. Similarly, an expert who accepts the AI's answer to save effort gradually loses their advantage over the AI.
Dynamic models relax the assumption of fixed skill: \citet{huang2026} show that reliance driven by observed performance can lock people into low skill. Closest to our work, \citet{aouad2026} model dynamic skill and show that AI assistance can make workers less productive by saving the effort that leads to skill acquisition. Their workers, however, choose how much effort to exert rather than whether to accept the AI's answer, so it remains open how the choice that reliance research evaluates changes once skill evolves.
Once skill evolves, the answer depends on whether workers weigh only the current task or also how today's choice shapes the skill they bring to future tasks. This motivates our first research question:
\begin{quote}
\textbf{RQ1:} When do workers solve tasks independently instead of accepting an AI's answer, depending on their skill, the AI's accuracy, and their strategic behavior?
\end{quote}

Valuing future skill, however, does not ensure that workers acquire or retain it. Practice pays off for a worker only if the skill gained earns more on future tasks than the effort and accuracy given up today. Some skills an organization needs, such as overseeing the AI, may not raise the worker's own payoff at all. Organizations have long paid for such practice implicitly, because the tasks they assigned trained workers as a by-product \cite{billett2001,becker1964,acemoglu1999}. Once AI can complete these tasks, organizations must decide explicitly which independent work to keep, much as space agencies schedule exercise rather than leaving it to the crew. One lever is to pay workers for correct independent answers. \citet{holstein2026} propose such an incentive to offset the effort of solving, which restores appropriate reliance. It does not, however, induce learning. A worker who is less accurate than the AI gives up correct answers as well as effort when solving, so a payment that covers only effort leaves them accepting and never practicing. This leads to our second research question:

\begin{quote}
\textbf{RQ2:} How can incentives
encourage workers to solve tasks independently and thereby acquire and maintain their skills?
\end{quote}

To answer these questions, we extend the accept-or-solve model of \citet{bansal2021} and \citet{holstein2026} from a single task to a sequence of tasks. We adopt the skill model of \citet{aouad2026}, in which skill improves through independent work and decays at a constant rate, but replace their choice of how much effort to exert with the accept-or-solve choice. A \emph{myopic} worker weighs only the effort and accuracy of the current task, whereas a \emph{strategic} worker also weighs today's effort against future skill. We then examine how incentives for correct independent answers can encourage workers to acquire and retain skill. Our work makes three contributions:
\begin{itemize}
  \item \textbf{Dynamic accept-or-solve model.} We extend the accept-or-solve framework with skill acquisition and decay and characterize how the value of future skill shapes reliance decisions. Strategic workers may solve independently even when the AI is more accurate, because doing so builds or preserves their skill.
  \item \textbf{Low-skill trap and expert deskilling.} We identify conditions under which reliance on AI leads to stagnating or eroding skill without a visible decline in task performance. Novices remain trapped at low skill levels, and experts lose proficiency while output accuracy stays at the AI's level.
  \item \textbf{Learning incentives.} We derive the minimum payment for correct independent answers that makes myopic workers acquire and retain skill. A higher skill target can reduce long-run incentive costs, either by making independent work self-sustaining or by lowering the payment needed to retain skill.
\end{itemize}
Together, these findings motivate an apprenticeship perspective on AI-assisted work, in which organizations pay for the practice that AI-assisted work no longer provides.

\section{Related Work}\label{sec:related}

Novices learn by taking part in actual work, starting with simple tasks and gradually taking on more central ones \cite{lave1991}. Experts support novices by demonstrating how tasks are done, giving feedback while they practice, and reducing their assistance as they become more skilled \cite{collins1989}. Organizations finance such practice when labor-market frictions allow them to capture part of its payoff \cite{becker1964,acemoglu1999}, and apprentices contribute through low wages and routine work that is valuable to their employers \cite{garicano2017,fudenberg2019}. Automation and AI weaken this link, as they restrict the effort through which workers learn \cite{beane2019}, limit the skill that novices develop when relying on tools they do not understand \cite{anthony2021}, change which skill is valued \cite{autor2025expertise}, and take over the entry-level work through which novices used to learn \cite{yu2026}.

Even where workers still perform the tasks themselves, AI assistance can lower their unaided skill: students who used AI without restriction performed worse once assistance was removed \cite{bastani2025}, as did learners who requested AI help during practice \cite{wu2026puzzles}. Professionals likewise lost skills after automation took over their tasks \cite{budzyn2025,rintakahila2023,natali2025}. These losses are consistent with shifting mental effort to external aids \cite{risko2016} and with skill decay through disuse \cite{arthur1998}, and echo the longstanding concern that automation causes skill decay needed to oversee it \cite{bainbridge1983}. Whether assistance harms learning depends on its design and use: a guided AI tutor avoided the harm of unrestricted access \cite{bastani2025}, and contrastive explanations improved decision-making skills \cite{bucinca2025}. Greater independent problem-solving effort predicted greater skill acquisition \cite{wu2026puzzles}, consistent with evidence that effortful practice and initial failure improve learning \cite{bjork1994,kapur2008}.

Whereas these studies examine how AI assistance affects learning, research on appropriate reliance asks when people should accept the AI's answer on a given task, based on human and AI accuracy \cite{leesee2004,schemmer2023,guo2024}, and how cognitive forcing functions can reduce overreliance on AI advice \cite{bucinca2021}. Some studies in this line also measure learning alongside accuracy \cite{gajos2022,bucinca2025, schemmer2023towards} or tailor AI assistance to support learning \cite{bucinca2026}. They vary how the AI assists, but do not model how workers trade off today's accuracy against future skill. Formal models capture the worker's choice between accepting the AI's answer and solving independently. \citet{bansal2021} formalize this choice as utility maximization, which \citet{holstein2026} extend with effort costs. They show that accepting can then be rational even when the worker is more accurate, and derive an incentive for correct independent answers that offsets this effort-induced overreliance. Both models, however, evaluate a single task with fixed skill.

Dynamic models relax this assumption of fixed skill. In the model of \citet{huang2026}, workers acquire skill through practice, lose it through disuse, and rely on AI based on its observed performance. As AI becomes more capable, such reliance locks people into low skill from a wider range of starting levels. Closest to our work, \citet{aouad2026} model skill as a birth--death process with effort-dependent learning and exogenous decay, and show that AI assistance can undermine long-run skill and productivity by reducing effort, even among strategic workers. We differ from \citet{aouad2026} in three ways. First, both learning and decay depend on the worker's choice on each task, so solving can let the worker acquire skill, whereas accepting the AI's answer can let skill decay. Second, because we retain the accept-or-solve model of \citet{bansal2021,holstein2026}, we can compare dynamic choices with appropriate reliance. Third, whereas apprenticeship models assume that routine work pays for practice \cite{fudenberg2019}, we derive the minimum incentive an organization must pay for solving independently to be rational, distinguishing temporary support to acquire skill from recurring support to retain it. We thereby extend to human--AI collaboration the insight of fairness research that single-step evaluations miss how decisions shape people over time \cite{damour2020,liu2020}.

\section{A Dynamic Accept-or-Solve Model}\label{sec:dynamic-model}

We extend the existing accept-or-solve model of \citet{bansal2021} from a single task to a sequence of tasks, where a worker's skills vary based on their previous decisions. In the following, we describe how we model both skill changes and the utility of a single task.

\subsection{Skill Dynamics}\label{sec:skill}
\begin{table}[!b]
\caption{One task in the dynamic accept-or-solve model. Utilities by choice and correctness of the final answer follow \citet[Table~1]{bansal2021}; the last column is this paper's extension. Skill otherwise stays at $p$, with $p^-=p$ at the lowest and $p^+=p$ at the highest level.}
\label{tab:utility}
\small
\begin{tabular}{@{}lccccl@{}}
\toprule
& & \multicolumn{2}{c}{Utility if final answer is} & & \\
\cmidrule(lr){3-4}
Choice & Correct with prob. & correct & incorrect & Expected utility & Skill for the next task\\
\midrule
Accept AI & $P_{AI}$ & $\gamma$ & $0$ & $u_A=\gamma P_{AI}$ & falls to $p^-$ with probability $\phi_d$\\
Solve independently & $p$ & $\gamma-\lambda$ & $-\lambda$ & $u_S(p)=\gamma p-\lambda$ & rises to $p^+$ with probability $\phi_u$\\
\bottomrule
\end{tabular}
\end{table}

Consider a sequence of tasks $t=0,1,2,\ldots$ from one \emph{task family}, such as reviewing code changes for bugs. Let $p_t$ denote the worker's \emph{unaided accuracy}, the probability of answering task $t$ correctly without AI assistance. We represent skill using a finite set of ordered levels:
\begin{equation}\label{eq:skill-grid}
 \mathcal P=\{a_1<a_2<\cdots<a_N\}\subset(0,1].
\end{equation}
For any skill level $p=a_i\in\mathcal P$, let $p^+=a_{i+1}$ and $p^-=a_{i-1}$ denote the adjacent higher and lower levels, respectively, with $p^+=p$ at the highest level and $p^-=p$ at the lowest.

When \emph{accepting AI}, the worker submits the AI's answer without spending effort. This answer is correct with probability $P_{AI}$, but skill goes unused and decays by one level to $p^-$ with probability $\phi_d\in(0,1]$; otherwise, it stays at $p$. When \emph{solving independently}, the worker instead spends effort $\lambda>0$ and submits their own answer, which is correct with probability $p$. In return for this effort, the worker acquires skill and rises one level to $p^+$ with probability $\phi_u\in(0,1]$; otherwise, skill stays at $p$. Table~\ref{tab:utility} summarizes both choices.

Learning depends on the effort of practice rather than on a correct answer, so even an incorrect independent solve can lead to skill acquisition \cite{corbett1994,kapur2008}. In line with \citet{corbett1994}, we assume $\phi_u$ and $\phi_d$ to be constant across skill levels within a task family. As in \citet{bansal2021}, the worker knows their own skill level $p$ and the AI's accuracy $P_{AI}$. Appendix~\ref{app:skill-dynamics} motivates this Markov chain and visualizes its dynamics.

Formally, solving instead of accepting thus raises the worker's expected skill level on the next task by
\begin{equation}\label{eq:next-skill-gap}
\mathbb E[p_{t+1}\mid p_t=p,\mathrm{Solve}]-\mathbb E[p_{t+1}\mid p_t=p,\mathrm{Accept}]
=\phi_u(p^+-p)+\phi_d(p-p^-)\geq0.
\end{equation}
The first term is the skill the worker can acquire, and the second is the skill loss that solving avoids. Appendix~\ref{app:skill-dynamics} derives Eq.~\eqref{eq:next-skill-gap} and the expected number of tasks until skill rises by one level.

\subsection{Single-Task Utility and Running Example}\label{sec:today}

As in \citet{bansal2021}, the worker's utility depends on the choice and on whether the final answer is correct (see Table~\ref{tab:utility}). A correct final answer earns $\gamma>0$, an incorrect one earns zero, and solving oneself requires the effort cost $\lambda$. While \citet{bansal2021} reward a correct answer with $\gamma$, they also penalize an incorrect one with $\beta$. Let $Y_t\in\{0,1\}$ indicate whether the final answer on task $t$ is correct, and let $U_t$ denote the worker's \emph{realized utility from that task}. At current skill $p$, the expected utilities of accepting and of solving are
\begin{equation}\label{eq:payoffs-dynamic}
 u_A=\gamma P_{AI},\qquad u_S(p)=\gamma p-\lambda.
\end{equation}

The \emph{immediate payoff gap} therefore is the difference between the expected payoff from accepting AI and the expected payoff from solving independently:

\begin{equation}\label{eq:immediate-payoff-gap}
 \Delta u(p)=u_A-u_S(p)
 =\underbrace{\lambda}_{\text{effort cost}}
 +\underbrace{\gamma(P_{AI}-p)}_{\text{accuracy gap}}.
\end{equation}
A positive gap means accepting the AI's answer immediately pays off compared to solving independently, whereas a negative gap means solving oneself pays off more. %

\begin{takeaway}{Each choice of a worker affects both today's payoff and tomorrow's skill}
For each task, the worker can either accept the AI's answer or solve the task independently. Accepting saves effort and yields a higher payoff today, unless the worker's accuracy advantage over the AI outweighs the effort cost. However, only by solving independently can the worker acquire skill and prevent its decay. Each choice therefore weighs today's payoff against the worker's future skill.
\end{takeaway}

\textbf{Running example.} \textit{To illustrate the model, imagine a software developer who reviews code changes for bugs with an AI assistant. Without help, Anna, a novice, finds the bug with 40\% probability; Sam, an expert, finds the bug with 80\% probability. Assume that skill ranges over $\mathcal P=\{.40,.50,.60,.70,.80\}$. With $\phi_u=.02$ and $\phi_d=.01$, a developer rises one level after $1/\phi_u=50$ solved tasks on average and falls one level after $1/\phi_d=100$ accepted tasks. Each bug found earns $\gamma=1$, and reviewing a change without the AI costs effort $\lambda=.05$. Anna and Sam work with one of two AIs. The \emph{competent AI} ($P_{AI}=.62$) is more accurate than Anna, but Sam's accuracy advantage over it ($.18$) exceeds the effort cost. Sam is also more accurate than the \emph{near-expert AI} ($P_{AI}=.77$), but his advantage ($.03$) falls short of the effort cost. Appendix~\ref{app:robustness} shows that our illustrative results hold across a range of parameter values.}

\section{Myopic and Strategic Workers}\label{sec:long-horizon}

To answer our first research question, we compare two workers who differ only in how strategic they are when solving tasks independently. A \textit{myopic worker} maximizes the utility of the current task, whereas a \textit{strategic worker} maximizes the discounted sum of utility over current and future tasks, and therefore values the skill that today's choice builds or preserves. For each worker, we ask how appropriate reliance would evaluate their choices.

\subsection{The Myopic Worker}\label{sec:myopic}

Research on human--AI decision-making mostly assesses reliance by the accuracy of the current answer alone \cite{schemmer2023,guo2024}. We apply appropriate reliance at the level of average accuracy, i.e., we compare the worker's skill level, which indicates the probability of solving correctly, with the AI's accuracy. Under appropriate reliance, a worker solves if $p\geq P_{AI}$ and accepts otherwise. Accepting when $p>P_{AI}$ is \emph{overreliance}, and solving when $p<P_{AI}$ is \emph{underreliance}.

A myopic worker also weighs the effort of solving. Maximizing only today's expected utility, the worker solves if $\Delta u(p)\leq0$, or equivalently
\begin{equation}\label{eq:myopic-rule}
 p\geq P_{AI}+\frac{\lambda}{\gamma}.
\end{equation}

The myopic threshold exceeds that of appropriate reliance by $\lambda/\gamma$. An expert who beats the AI by less than this margin therefore still accepts, which is the effort-induced overreliance described by \citet{holstein2026}. Yet, neither appropriate reliance nor the myopic worker considers that the same choice also shapes tomorrow's skill~\eqref{eq:next-skill-gap}.

\textbf{Running example.} \textit{With the competent AI, the myopic threshold is $.62+.05=.67$. Anna ($.40$) therefore accepts, and only developers at $.70$ or above solve. With the near-expert AI, the threshold rises to $.82$, above even Sam. Although Sam ($.80$) is more accurate than the AI, Sam faces $\Delta u=.05+(.77-.80)=.02>0$ and accepts.}

\subsection{The Strategic Worker}\label{sec:strategic}

A strategic worker recognizes that solving a task today can raise or preserve their skill and thereby their utility on future tasks. To weigh today's utility against future utility, the worker discounts the utility of the task $t$ steps ahead by $\delta^t$, with a discount factor $\delta\in[0,1)$. This discount factor is the probability that the worker expects to perform another task of the same family after the current one, and it does not depend on whether the worker accepts or solves. The weight $\delta^t$ is then the probability that the worker still performs these tasks $t$ steps ahead, and the worker expects to perform $H=1/(1-\delta)$ tasks, including the current one (Appendix~\ref{app:delta}). Although the sequence of tasks itself does not end, the worker may expect to leave it, for example by moving to another role. A myopic worker has $\delta=0$ and thus $H=1$, whereas a strategic worker has $\delta>0$ and thus $H>1$.

Let a \emph{policy} $\pi$ specify the worker's accept-or-solve choice at each skill level, and define the \emph{lifetime value} $V(p)$ as the greatest expected discounted utility starting from skill $p$:
\begin{equation}\label{eq:dynamic-value}
 V(p)=\max_{\pi}\;\mathbb E_{\pi}\!\left[
  \sum_{t=0}^{\infty}\delta^t U_t
  \;\middle|\; p_0=p\right].
\end{equation}
By the principle of optimality, the worker chooses on each task whichever option yields the higher sum of current utility and discounted lifetime value from the next task on. This choice satisfies the Bellman equation
\begin{equation}\label{eq:dynamic-bellman}
\begin{aligned}
 V(p)=\max\big\{
 &u_A+\delta[(1-\phi_d)V(p)+\phi_d V(p^-)],\\
 &u_S(p)+\delta[(1-\phi_u)V(p)+\phi_u V(p^+)]
 \big\}.
\end{aligned}
\end{equation}
The first expression is the value of accepting AI now, including possible skill decay. The second is the value of solving now, including possible learning. Comparing the two expressions gives the strategic worker's decision rule:
\begin{equation}\label{eq:future-premium}
\underbrace{\Delta u(p)}_{\text{immediate payoff gap}}
\;\leq\;
\underbrace{\delta \left[ \phi_u \left( V(p^+) - V(p) \right) + \phi_d \left( V(p) - V(p^-) \right) \right]}_{\text{future value of solving}}.
\end{equation}
The right-hand side captures the future payoff of possible learning \emph{and} of avoided skill decay. When $\delta=0$, it vanishes, and Eq.~\eqref{eq:future-premium} reduces to the myopic rule in Eq.~\eqref{eq:myopic-rule}. Appendix~\ref{app:bellman} derives Eq.~\eqref{eq:dynamic-bellman}, and Appendix~\ref{app:solution} explains how we solve it exactly on the finite skill grid.

\begin{figure}[tb]
    \centering
    \includegraphics[width=1\linewidth]{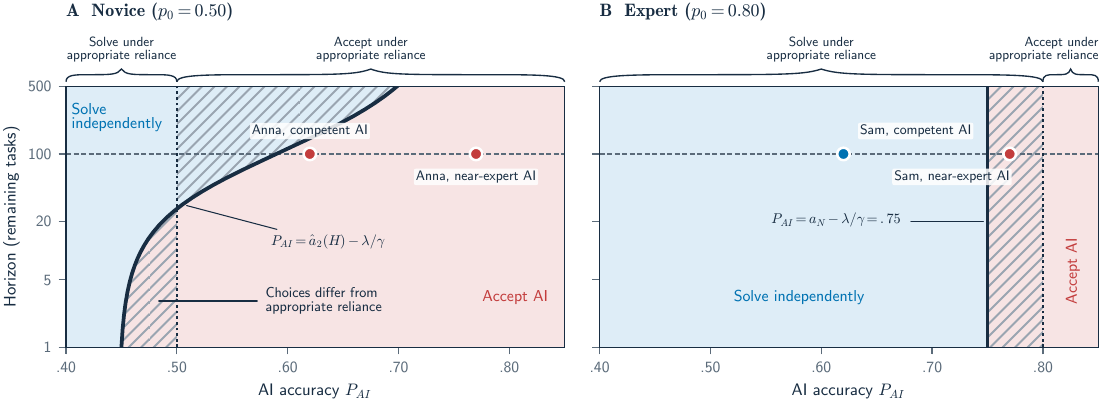}
\caption{Rational choices for a strategic worker at AI accuracy $\PAI$ and expected horizon $H$. Blue shading indicates solving independently, and red shading indicates accepting the AI's answer. The braces show the appropriate-reliance benchmark, which favors solving when $p_0 \geq \PAI$. Hatching marks where the strategic choice differs from that benchmark. Points on the dashed line show the running examples at $H=100$. \textbf{(A)} The novice ($p_0=.50$) accepts the competent AI at $H=100$ despite the opportunity to build skill. \textbf{(B)} The expert ($p_0=.80$) accepts the near-expert AI at every horizon. Appendix~\ref{app:closed-form} derives the strategic boundaries.}
    \label{fig:who-solves}
\end{figure}

Because of this future payoff, a strategic worker may solve independently even when the AI is more accurate ($p<\PAI$). Figure~\ref{fig:who-solves} compares these choices with appropriate reliance in the running example. Whether a worker starts to solve depends on the horizon and the learning probability only through $\phi_u(H-1)$, the number of skill levels the worker can expect to acquire over the remaining tasks. The decay probability plays no role in this decision, as a worker who keeps solving never loses skill (Appendix~\ref{app:rob-checks}).

\textbf{Running example.} \textit{Let $\delta=.99$, so Anna and Sam each expect 100 remaining tasks. At $.60$ with the competent AI, Anna faces $\Delta u=.07$ and still solves, whereas a myopic Anna accepts and risks falling back to $.50$. Solving first costs Anna expected utility, a shortfall that peaks at $1.30$ after 38 tasks and is recouped after 80 tasks. Over all remaining tasks, the discounted gain relative to always accepting is $4.06$ (Appendix~\ref{app:strategic-return})}.

\begin{takeaway}{Strategic behavior can make solving worthwhile even when the AI is more accurate}
A myopic worker maximizes the payoff of the current task and solves only if their accuracy advantage over the AI outweighs the effort cost. Such a worker may therefore accept even when the AI is less accurate. A strategic worker also values skill for future tasks and may solve even when the AI is more accurate because the acquired skill may increase future payoffs.
\end{takeaway}

\subsection{The Low-Skill Trap and Expert Deskilling}\label{sec:foresight-fails}

Valuing future skill, however, does not guarantee that a worker will acquire or retain skill. Even for a strategic worker, acquiring skill may cost more than its future payoff justifies---a situation we call the \emph{low-skill trap}. Retaining existing skill may likewise offer no payoff advantage over accepting AI, a situation we term \emph{expert deskilling}.

The \emph{low-skill trap} arises when learning does not pay off. To acquire skill a novice must solve independently, forgoing immediate payoff from accepting the AI's answer, until reaching a skill level at which solving pays on its own. The lifetime value in Eq.~\eqref{eq:future-premium} accounts for the entire horizon $H$ with their opportunities for acquiring skill: the worker weighs the benefits of potentially becoming an expert. However, at a low skill level and a short horizon, the discounted utility can be smaller than the accumulated cost of reaching a skill level through the effort of solving independently. Put simply, solving independently then does not pay off -- the worker chooses to accept the AI's answers and skill can only retain or decay, leaving the worker trapped at a low skill level. Appendix~\ref{app:obstacles} states the condition under which the trap arises.

\emph{Expert deskilling} arises when retaining existing skill offers no payoff advantage over accepting AI. At the highest skill level, further learning is impossible, so only the retention term in Eq.~\eqref{eq:future-premium} remains. That retention of skill is only valuable to the worker if it improves future payoffs. Suppose $a_N < P_{AI}+\lambda/\gamma$. Then even experts at the highest skill level earn less in expectation by solving, $\gamma a_N-\lambda$, than by accepting, $\gamma P_{AI}$, and every lower skill level earns less still. Accepting on every task is then rational regardless of starting skill, so $V(p)=\gamma P_{AI}/(1-\delta)$ for every $p\in\mathcal P$. Both future-value terms are zero, and strategic behavior cannot make retention worthwhile.

The low-skill trap and expert deskilling hold across all examined parameter values  (Appendix~\ref{app:robustness}). For every finite horizon, there is a range of AI accuracies in which independent work would pay at the top level, yet a novice accepts. Longer horizons and faster learning narrow this range and shift it toward more accurate AIs, but cannot remove it. Expert deskilling starts as $\PAI>a_N-\lambda/\gamma$, whatever the horizon, learning or decay.

\textbf{Running example.} \textit{With the competent AI, Anna at $.50$ accepts. Reaching $.60$ has some utility, but not enough to cover the gap of $.17$. At $.40$, even a step to $.50$ leaves Anna where accepting remains rational, so $V(.40)=V(.50)$ and this step is worthless. Anna thus stays at $.40$ or slides there from $.50$, while output accuracy stays at $.62$. Since Anna is less accurate than the AI, appropriate reliance scores every acceptance as appropriate. With the near-expert AI, Sam beats the AI by $.03$, less than the effort cost of $.05$. Sam accepts on every task and falls one level after 100 tasks on average, while output accuracy stays at $.77$. Appropriate reliance first labels these acceptances as overreliance, but as appropriate once Sam falls to $.70$.}

\begin{takeaway}{Strategic behavior alone does not prevent the low-skill trap or expert deskilling of workers} Even a strategic worker may fail to acquire or retain skill. In the \emph{low-skill trap}, a novice whose accuracy is far below that of the AI keeps accepting because the cost of practicing until independent work pays off exceeds the future payoff. In \emph{expert deskilling}, an expert whose accuracy advantage over the AI no longer outweighs the effort cost accepts the AI's answer on every task, regardless of how many future tasks. In both cases, output accuracy remains at the AI's level, while the worker's unaided accuracy declines. %
\end{takeaway}

\section{Dynamic Incentives for Independent Solving}\label{sec:dynamic-incentives}

Strategic behavior can leave a worker in the low-skill trap and let the skill of an expert decay despite a high skill level. To answer our second research question, we therefore investigate next how incentives, i.e., monetary payments, can make independent work worthwhile for workers. Following \citet{holstein2026}, we make three assumptions: first, we assume that there exists an organization that pays an incentive for correct independent answers; second, we assume that the organization can observe a worker's skill and verify that an answer was produced without the AI; third, we assume a myopic worker for now, because a payment that incentivizes a myopic worker also incentivizes a strategic one (Appendix~\ref{app:incentives}).

Let $\theta(p)\geq0$ be an incentive paid only for a correct independent answer at skill level $p$.
We resolve ties in favor of solving. A myopic worker therefore solves if and only if the expected incentive covers the immediate payoff gap:
\begin{equation}\label{eq:incentive-ic}
 p\theta(p)\geq\Delta u(p).
\end{equation}

\subsection{Why the Effort Incentive is Insufficient}\label{sec:effort-incentive}

\citet{holstein2026} pay $\lambda/p$ for a correct independent answer in a model with fixed skill. In our dynamic setting, we evaluate this \emph{effort incentive} at the worker's current skill, $\theta^{\mathrm{eff}}(p_t)=\lambda/p_t$. At current skill $p$, this changes the expected utility of solving to
\begin{equation}\label{eq:effort-incentive-payoff}
 u_S^{\mathrm{eff}}(p)
 =\gamma p-\lambda+p\frac{\lambda}{p}
 =\gamma p.
\end{equation}
Put simply, the incentive payment changes the expected utility to become only the expected earnings. Since the utility from accepting the AI's answer is $\gamma P_{AI}$ (Eq.~\ref{eq:payoffs-dynamic}), a myopic worker therefore solves if and only if $p\geq P_{AI}$. The effort incentive thus restores appropriate reliance, as \citet{holstein2026} intend, as workers no longer accept a less accurate AI merely to avoid effort.

Nevertheless, when workers accept the AI's answer, they forgo the opportunities to acquire skill. So while the effort incentive restores appropriate reliance, it may still lead to skill decay. Strategic behavior does not necessarily tackle that issue: when the effort incentive is paid, a strategic worker faces the low-skill trap and expert deskilling (Appendix~\ref{app:rob-checks}).

\begin{figure}[tb]
    \centering
    \includegraphics[width=1\linewidth]{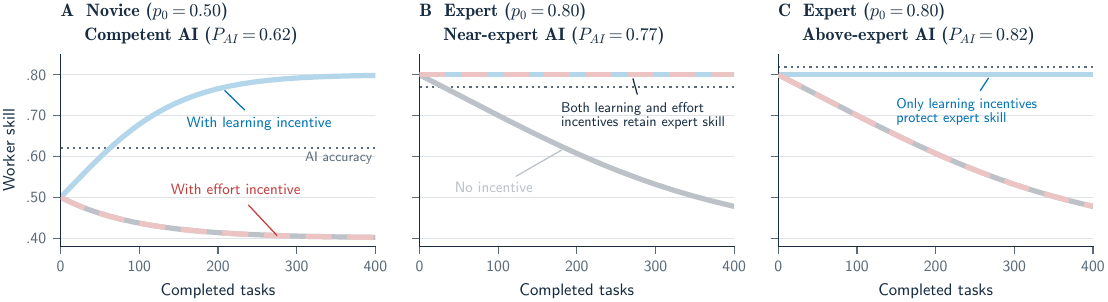}
    \caption{Expected worker skill over completed tasks for a myopic worker with no incentive (grey), the effort incentive of \citet{holstein2026} (red), or the learning incentive in Eq.~\eqref{eq:minimum-current-incentive} (blue). The dotted horizontal line marks AI accuracy. \textbf{(A)} With competent AI, only the learning incentive builds novice skill. \textbf{(B)} With near-expert AI, both incentives retain expert skill at $.80$, so their curves overlap. \textbf{(C)} With above-expert AI, only the learning incentive retains expert skill. In panels A and C, the effort and no-incentive curves overlap.}
    \label{fig:incentive-mechanisms}
\end{figure}

\textbf{Running example.} \textit{With the competent AI, Anna accepts at $.40$, $.50$ and $.60$ despite the effort incentive, because each level is below $.62$. Every acceptance is appropriate, but Anna stays in the low-skill trap (Figure~\ref{fig:incentive-mechanisms}A). With the near-expert AI, the effort incentive keeps Sam solving at $.80$ and thus prevents deskilling (Figure~\ref{fig:incentive-mechanisms}B). However, once the AI becomes more accurate than Sam, accepting pays more at every skill level, even with the incentive, and Sam deskills again (Figure~\ref{fig:incentive-mechanisms}C). The effort incentive thus raises the AI accuracy at which experts start to deskill from $a_N-\lambda/\gamma$ to $a_N$, but no further.}

\begin{takeaway}{The incentive paid for the effort restores appropriate reliance but not learning}
The incentive proposed by \citet{holstein2026} pays the worker for each correct independent answer so that the effort cost is covered in expectation. A myopic worker then solves if and only if they are at least as accurate as the AI, so effort no longer causes overreliance. However, a worker who is less accurate than the AI never starts to practice, and experts deskill as soon as the AI becomes more accurate than they are.
\end{takeaway}

\subsection{A Learning Incentive}\label{sec:learning-incentive}

To make solving worthwhile below the AI's accuracy, an incentive must therefore cover both the effort and the difference between the AI's accuracy and the worker's skill level. Since both enter the immediate payoff gap $\Delta u(p)$, solving~Eq.~\eqref{eq:incentive-ic} for $\theta(p)$ yields the minimum nonnegative learning incentive with $[x]^+=\max\{x,0\}$:
\begin{equation}\label{eq:minimum-current-incentive}
 \theta^{\min}(p)
 =\frac{[\Delta u(p)]^+}{p}
 =\frac{[\lambda+\gamma(P_{AI}-p)]^+}{p}.
\end{equation}
Wherever $\Delta u(p)\geq0$, this incentive splits into the effort incentive and an adjustment for the accuracy gap:
\begin{equation}\label{eq:incentive-split}
 \theta^{\min}(p)
 =\underbrace{\frac{\lambda}{p}}_{\text{effort incentive}}
 +\underbrace{\frac{\gamma(P_{AI}-p)}{p}}_{\text{accuracy gap}}.
\end{equation}

Below the AI's accuracy, the second term compensates for the correct answers the worker forgoes by solving, so that the worker practices and can acquire skill. Above it, the worker's own accuracy advantage covers part of the effort, so the second term is negative and the incentive falls below the effort incentive. The learning incentive therefore shrinks as more skills are acquired and vanishes once the accuracy advantage outweighs the effort, which also spares the organization from paying workers who would solve anyway. Appendix~\ref{app:incentives} derives Eqs.~\eqref{eq:incentive-ic}--\eqref{eq:incentive-split}.

\textbf{Running example.} \textit{With the competent AI, Anna needs an expected incentive of $.27$ per task at $.40$, $.17$ at $.50$, and $.07$ at $.60$, of which $.05$ covers effort and the rest the accuracy gap. At $.70$, solving pays without an incentive. With the near-expert AI, Sam at $.80$ needs only $.02$ instead of the effort incentive's $.05$.}

\begin{takeaway}{The learning incentive compensates for both the effort and the accuracy gap of the worker to the AI}
To make independent work worthwhile for a worker who is less accurate than the AI, the organization must compensate the worker for both the effort and the correct answers the worker forgoes by not accepting the AI's answer. This learning incentive is highest for novices and decreases as the worker acquires skill. For a worker who is more accurate than the AI, the learning incentive is lower than the effort incentive, and it ends at the skill level at which independent work is rational without an incentive.
\end{takeaway}

\subsection{Target Skill Levels}\label{sec:targets}

Paying the learning incentive at every skill level would also pay for practice the organization may not need. We therefore let the organization choose a \emph{target skill level} $p^\star\in\mathcal P$, which can stand for the foundational skill that more advanced work requires. The organization requires independent solving at or below the target and leaves choices above it unsupported. The minimum incentive per correct independent answer that implements this policy for myopic workers is
\begin{equation}\label{eq:incentive-target}
 \theta(p;p^\star)=
 \begin{cases}
 \dfrac{[\lambda+\gamma(P_{AI}-p)]^+}{p}, & p\leq p^\star,\\[5pt]
 0, & p>p^\star.
 \end{cases}
\end{equation}
For a novice below the target, the target is a goal to acquire. For an expert at or above it, the target is a floor to retain. Whether payments end depends on whether the induced policy eventually reaches levels at which solving continues without support.

\textit{Acquiring a target.} Below the target, the incentive makes the worker solve at every level, so the worker can acquire skill, but skill never decays. With $\phi_u>0$, the worker therefore reaches the target almost surely, after $r/\phi_u$ independent tasks on average when starting $r$ levels below it. Reaching the target does not by itself end the payments, because the worker accepts again just above it. Payments end for good only once the worker reaches a level at which solving pays without an incentive. If that level lies one step above the target, the incentive keeps the worker practicing at the target until the step occurs, and the worker then keeps solving on their own. If that level lies further above, the worker accepts just above the target, slides back and triggers the incentive anew, so payments recur indefinitely. Payments thus end if and only if solving pays without an incentive one level above the target (Appendix~\ref{app:incentives}).

\textbf{Running example.} \textit{With the competent AI, a target of $.60$ bridges Anna to independence. The incentive supports practice up to $.60$, and Anna eventually reaches $.70$, where solving pays without an incentive ($.70\geq.67$). Starting from $.40$, the expected total spending is $25.50$, after which payments stop (Appendix~\ref{app:incentives}). A target of $.50$ falls short. After reaching $.60$, Anna accepts again, falls back to $.50$ and triggers the incentive anew, so payments recur indefinitely. A higher target can thus turn recurring payments into temporary support, although it need not cost less over every finite horizon.}

\textit{Retaining a target.} Keeping skill at the target requires continued practice, but not necessarily continued payments. Whenever skill sits at the target, the incentive makes the worker solve. Since solving never lowers skill and accepting lowers it by at most one level, skill cannot fall below $p^\star$ while the incentive is in place. Payments continue only where solving does not pay on its own. At the highest target, $p^\star=a_N$, no further learning is possible, so if $\Delta u(a_N)>0$, the worker needs a positive expected incentive on every task for as long as the skill is to be kept.

A lower target lets skill decay before the incentive restarts practice, but it can cost more in the long run. The incentive then pays repeatedly to acquire skill that was allowed to decay, and the larger payoff gap at the lower level makes each supported task more expensive. Appendix~\ref{app:incentives} gives the condition under which the higher target costs less.

\textbf{Running example.} \textit{With the near-expert AI, keeping Sam at $.80$ costs $.02$ per task in expectation, still paid only for correct independent answers. With a target of $.70$, Sam accepts until falling to $.70$, practices with the incentive until returning to $.80$ and then accepts again. Each supported task at $.70$ costs $.12$ instead of $.02$, and the long-run payment is $.04$ per task. Here, the higher target thus retains more skill at half the long-run cost.}

\begin{takeaway}{Incentives for acquiring skill can be temporary, whereas retaining skill can require continuous incentives} Support to acquire a skill target ends once the worker reaches a level at which independent work is rational without incentive; otherwise skill repeatedly decays and incentives need to recur. Thus, retaining a skill level at which independent work is not rational requires incentives for as long as the skill is needed.
\end{takeaway}

\section{Discussion}\label{sec:discussion}

Our primary theoretical contribution is to extend the existing accept-or-solve framework \cite{bansal2021,holstein2026} from a single task to a sequence of tasks in which each choice changes the skill of a worker. While research on AI-assisted decision-making has evaluated reliance within a task \cite{schemmer2023,guo2024}, we show that the same choices determine the skill workers bring to future tasks. A myopic worker solves only when their accuracy advantage over the AI outweighs the effort, whereas a strategic worker may solve even when the AI is more accurate, a choice that appropriate reliance classifies as underreliance (RQ1).  In line with prior dynamic models \cite{aouad2026,huang2026}, we show that strategic behavior does not prevent the low-skill trap or expert deskilling. In both cases, output accuracy remains at the level of the AI while skill decays. This may explain the empirically observed deskilling \cite{budzyn2025,shen2026}: when workers are rewarded for correct answers, accepting AI answers can be rational even though skill decays, and the decline can go unnoticed \cite{macnamara2024}. Since such effects unfold over many tasks and are hard to observe in short experiments, our model makes them analyzable before longitudinal evidence exists.

Our incentive analysis subsequently shows that paying for correct independent solving does not by itself build skill (RQ2). The effort incentive of \citet{holstein2026} restores appropriate reliance but not learning. The learning incentive, by contrast, additionally compensates for the accuracy gap between the worker and the AI, so that novices start to practice and experts retain their skill. Support to acquire a target skill level can end once independent solving pays without an incentive, whereas retaining a skill level at which independent work does not pay requires continuous incentives. This shows that incentive design for AI-assisted work must target the future skill of the worker rather than the accuracy of the current answer alone.

\subsection{Appropriate Reliance and Skill}\label{sec:impl-researchers}

Since appropriate reliance compares only current accuracies, it labels a novice's acceptances in the low-skill trap as appropriate and a strategic worker's practice below the AI's accuracy as underreliance. Once an expert falls below the AI's accuracy, it also labels the expert's remaining skill decay as appropriate.  As \citet{damour2020} showed for fairness, short-term evaluations of workers' accept-or-solve choices miss long-run consequences. We therefore call for extending appropriate reliance to account for how a decision changes a worker's future skill.

Empirical studies of human--AI decision-making mostly measure output accuracy \cite{lai2023}. As our model shows, such studies can report stable output accuracy while the unaided accuracy that workers need for oversight and advanced work decays unnoticed. Evaluations of human--AI collaboration should therefore state their horizon and track unaided accuracy alongside output accuracy.

\subsection{Incentives for Learning}\label{sec:impl-organizations}

With AI assistance, skill acquisition is becoming an explicit cost for organizations, which is new since acquiring skill has largely occurred as a by-product of assigned tasks \cite{billett2001}. When AI is used for these tasks, as in robotic surgery \cite{beane2019}, this by-product disappears. Medical residency illustrates an alternative: residents are paid to do the work through which they learn, and their slower work is accepted as the price of future skill. Organizations can transfer this model as an AI-era apprenticeship that pays for the practice AI-assisted work no longer provides. Since the expected payment to acquire a target falls as the learning probability rises (Eq.~\eqref{eq:app-acquisition-cost}), feedback and well-chosen practice tasks that speed learning also reduce the incentive an organization must pay. Since retaining a higher-skill target can cost less than repeatedly restoring a lower one, organizations may need to budget differently and compare targets by long-run spending. Strategic workers never cost more than myopic workers under the same target, so organizations that help workers value skill on future tasks may also pay less.

Since the learning incentive compensates for both the effort and the reward forgone by not accepting AI answers (Eq.~\eqref{eq:minimum-current-incentive}), an activity that lets workers acquire skill while forgoing less reward would need a smaller incentive. Evidence on whether such AI-supported activities lead to skill acquisition is mixed: answering before seeing the AI's answer improved immediate decisions but not learning \cite{gajos2022}, whereas explanations that contrast the AI's answer with a predicted human choice improved learning more than explanations that only justify the AI's answer \cite{bucinca2025}. Researchers should therefore identify which activities lead to skill acquisition at the lowest forgone reward. This includes practice whose answer does not replace the AI's answer, such as simulated cases or independent attempts submitted only for review. Such practice forgoes no reward on the current task, so its incentive would need to cover only the effort cost.

While a worker is less accurate than the AI, promotion rankings based on output accuracy favor colleagues who accept the AI's answer. The organization then pays for learning through one part of its reward system and penalizes it through another. Protected learning assignments and appraisals that recognize skill acquisition could ease this conflict. Whatever construct an appraisal measures, whether output accuracy, unaided accuracy, or professional judgment, workers will optimize for it \cite{jacobs2021measurement}.

\subsection{Governing Skill Acquisition and Retention}\label{sec:impl-policymakers}

Effective oversight under the EU AI Act requires adequate epistemic access \cite{sterz2024}. Such policies require and presume that workers can perform the tasks \cite{green2022}. Yet, our model shows that the skill on which oversight depends can decay while output accuracy stays stable. This suggests that oversight mandates should cover retaining the skill of overseers, not only assigning oversight. The EU AI Act already requires deployers of high-risk AI systems to provide human overseers with the necessary competence, training, and support \cite[Art.~26(2)]{euaiact2024}, but leaves open how deployers retain this skill once AI-assisted work no longer exercises it. Our learning incentive offers one mechanism for doing so, complementing automation research, which has long argued that workers, such as pilots, must regularly practice manual control, on the job or in simulators, to retain the skill needed to take over from automation \cite{bainbridge1983}.

Workers carry transferable skills to future employers, so employers may not capture the full returns of training, although labor-market frictions can let them benefit from investments in general skills \cite{acemoglu1999}. Conversely, organizations can offer opportunities for skill acquisition to retain workers, since workers may stay where they can keep acquiring skill. Worker mobility can also lead organizations to offer skill-building opportunities mainly to workers who are already skilled \cite{singh2026}. Apprenticeships, sectoral training funds, and public co-funding can share these costs without restricting mobility or leaving less-skilled workers behind.

Existing institutions offer a precedent for governing such investments. German works councils, for example, co-determine in-company training when employer measures leave the skills of employees insufficient for their changed work, and they co-determine the introduction of technical devices that monitor conduct or performance \cite[\S\S~87(1) No.~6, 97(2)]{betrvg}. Implementations of our mechanism engage both rights, since they require assessing skill and verifying that answers were produced without the AI.

\subsection{Fairness, Accountability, and Power}\label{sec:ethics}\label{sec:normative}

The learning incentive reverses a familiar pattern: worker pay usually rises with skill, whereas the learning incentive is largest for the least skilled worker. Since the incentive does not reward lower skill but compensates for the effort and the reward a worker forgoes while practicing below the AI's accuracy. Without it, the starting skill of a worker determines whether learning pays at all, and novices remain in the low-skill trap not because they cannot learn but because learning is not rational. A larger incentive for novices thus equalizes the opportunity to acquire skill rather than the outcome, much as subsidizing the cost of investment can prevent disparate equilibria when individuals invest rationally in their qualifications \cite{liu2020}. Colleagues who receive less pay per task for the same work may nevertheless perceive the incentive as unfair, and an organization with a fixed budget has a calculable reason to support workers who are already skilled.

The costs of learning fall partly on people whom the incentive does not compensate. In our example of acquiring a target, a novice supported from .40 to .60 submits about 18 more incorrect answers in expectation than the AI would. The clients affected by these errors are not necessarily those who later benefit from the skill of the worker. Without support for retaining skill, workers in turn risk being held responsible for failures of systems over which they have limited control \cite{elish2019}.

In our model, an external party, like an organization, chooses the skill target, observes the skill, and sets the incentive. With this, they decide on which skills workers require or retain and up to which skill level. Letting a skill decay might be legitimate, but it becomes harmful when the interests of workers contradict those of the organization. Since workers contribute the effort, are potentially affected by their organization's skill assessments, and may value skills differently, they should participate in setting targets and deciding how assessments are used, for example through works councils. Skill assessments could inform discipline, and because the incentive is smaller at higher skill levels, workers may benefit from appearing less skilled, while assessments robust to such manipulation can impose their costs unequally \cite{hu2019,milli2019}.

\subsection{Limitations and Future Research}\label{sec:limitations}

Our numerical examples do not establish how often the low-skill trap or expert deskilling occurs, how long workplace learning takes, or which payments workers would require in real-world settings. Longitudinal studies should estimate effort costs, task rewards, learning and decay probabilities, and the horizons over which workers expect to benefit. They should also test whether incentives induce practice, whether that practice builds and retains the intended skill, and how payments and workload evolve. Since acquiring a target assumes indefinitely continuing tasks, our results do not guarantee reaching a target within a fixed-term employment period.

In our model, skill is acquired only through independent solving and represents unaided accuracy in one task family. If workers also acquire skill through engaged AI use, the low-skill trap may weaken, and the learning incentive could pay for skill acquisition that AI-assisted work would provide anyway. Retaining a target relies on the assumption that solving never decays skill and on accepting decays it by at most one skill level. Future research should consider learning and decay rates that vary across tasks and over time, as well as declines larger than one level. Models with several linked skills could examine whether practicing foundational skills lowers the cost of acquiring advanced ones. Such models could also capture collaboration with AI as a skill of its own that makes the original skill less relevant. In writing, for example, directing and editing drafts with AI assistance may matter more than drafting text unaided.

The minimum incentive applies to myopic workers paid for correct independent answers, with observable skill, verifiable independence, ties resolved in favor of solving, and known effort costs and task rewards, and workers are assumed to know their own accuracy. Since unsuccessful solving attempts might also teach, payments for attempts, protected practice time, and non-monetary incentives should be studied in the future.

Finally, we analyze a single worker and measure the organization only through incentive spending. This spending omits the accuracy forgone during practice and the value of the acquired skill, so our comparisons of targets do not identify a welfare-maximizing target. Extending the model to an organizational objective and to heterogeneous workers, whose learning probabilities and starting levels determine the cost of reaching a target, remains future work.

\section{Conclusion}\label{sec:conclusion}

Today's output accuracy provides an incomplete view of how AI-assisted work will shape tomorrow's workforce. We extended an established accept-or-solve model to a sequence of tasks in which solving independently can let the worker acquire skill, whereas accepting the AI's answer can let skill decay. Strategic behavior can make independent work worthwhile even when the AI is more accurate, although appropriate reliance classifies this choice as underreliance. However, strategic behavior alone does not guarantee that workers will acquire or retain skill. Novices can remain in a low-skill trap, and experts can deskill, while output accuracy remains at the AI's level. The effort incentive restores appropriate reliance, but not learning. The learning incentive we derive additionally compensates for the accuracy gap between the worker and the AI. Support to acquire a target skill level can end once independent work pays without an incentive, whereas retaining a skill level at which independent work does not pay requires continuous incentives.
Just as no astronaut can stay strong by waiting for gravity to return, no organization can keep its experts by waiting for AI to fail. Exercise must be scheduled while work is still being done, so that experts retain their skill and novices acquire it in the first place.
Space agencies only began scheduling exercises after astronauts returned to Earth too weak to stand. Organizations adopting AI cannot afford to learn this lesson the same way. The loss of skill remains hidden until the AI fails and someone must take over. Organizations that wait until then will find that the necessary skills are gone, and there is no one left to teach them.

\section*{Generative AI Usage Statement}

The authors used the following generative AI tools during manuscript preparation: Claude Opus 5.5 (Anthropic), GPT6.1-Sol and GPT6-Astra (OpenAI), DeepL Write, and Grammarly. We used these tools for literature search, drafting text and language editing. We also used them to write code for the model, robustness checks, figures, and to check the mathematical derivations. The authors reviewed and edited all AI-generated content and take full responsibility for the content of this paper.

\section*{Ethical Considerations and Adverse Impacts}

This work is theoretical and thus involves no human participants, no personal data, or deployed system. The numerical examples use illustrative parameter values, not empirical estimates.

\clearpage
\appendix
\section*{Appendix}
\definecolor{chainSolve}{HTML}{007C83}
\definecolor{chainAccept}{HTML}{AB552A}
\definecolor{chainInk}{HTML}{182D42}

\section{Derivations of the model equations}\label{app:derivations}

This appendix derives every equation in Sections~\ref{sec:dynamic-model}--\ref{sec:dynamic-incentives} step by step and shows how the numbers in the running example follow from them. Table~\ref{tab:app-notation} shows the notation, and Table~\ref{tab:app-guide} shows where each equation of the main text is derived in the appendix. All numbers are exact expectations computed from the transition rules, without simulation.

\begin{table}[h]
\caption{Notation and running-example values.}
\label{tab:app-notation}
\small
\begin{tabular}{@{}llL{4.2cm}@{}}
\toprule
Symbol & Meaning & Running example\\
\midrule
$\mathcal P=\{a_1<\dots<a_N\}$ & skill levels (unaided accuracy) & $\{.40,.50,.60,.70,.80\}$, $N=5$\\
$p,\ p^+,\ p^-$ & current level and its upper and lower neighbor & e.g.\ $p=.60$, $p^+=.70$, $p^-=.50$\\
$\PAI$ & accuracy of the AI & competent $.62$, near-expert $.77$, above-expert $.82$\\
$\gamma$ & reward for a correct final answer & $1$\\
$\lambda$ & effort cost of solving independently & $.05$\\
$\phi_u$ & probability of a learning step after solving & $.02$\\
$\phi_d$ & probability of a decay step after accepting & $.01$\\
$\delta$ & discount factor (probability that another task follows) & $.99$, so $H=1/(1-\delta)=100$\\
$\Delta u(p)$ & immediate payoff gap, Eq.~\eqref{eq:immediate-payoff-gap} & e.g.\ $\Delta u(.60)=.07$ with the competent AI\\
$V(p)$ & lifetime value, Eq.~\eqref{eq:dynamic-value} & ---\\
$\theta(p)$ & incentive per correct independent answer & ---\\
\bottomrule
\end{tabular}
\end{table}

\begin{table}[h]
\caption{Where each equation of the main text is derived.}
\label{tab:app-guide}
\small
\begin{tabular}{@{}lll@{}}
\toprule
Main text & Content & Appendix\\
\midrule
Eqs.~\eqref{eq:skill-grid}, \eqref{eq:next-skill-gap} & skill levels as a Markov chain, expected next-task accuracy & \ref{app:skill-dynamics}\\
Table~\ref{tab:utility}, Eqs.~\eqref{eq:payoffs-dynamic}, \eqref{eq:immediate-payoff-gap} & utility of one task, immediate payoff gap & \ref{app:one-task}\\
Eq.~\eqref{eq:myopic-rule} & myopic decision rule & \ref{app:myopic}\\
Eq.~\eqref{eq:dynamic-value} & lifetime value, meaning of $\delta$ & \ref{app:delta}\\
Eqs.~\eqref{eq:dynamic-bellman}, \eqref{eq:future-premium} & Bellman equation, strategic decision rule & \ref{app:bellman}\\
--- & how the model is solved & \ref{app:solution}\\
Figure~\ref{fig:who-solves} & closed form of the strategic choice, policy regions & \ref{app:closed-form}\\
Section~\ref{sec:foresight-fails} & low-skill trap, expert deskilling & \ref{app:obstacles}\\
Eqs.~\eqref{eq:incentive-ic}--\eqref{eq:incentive-target} & incentives and their costs & \ref{app:incentives}\\
Section~\ref{sec:strategic}, running example & strategic return of independent work & \ref{app:strategic-return}\\
\bottomrule
\end{tabular}
\end{table}

\subsection{Skill dynamics as a Markov chain}\label{app:skill-dynamics}

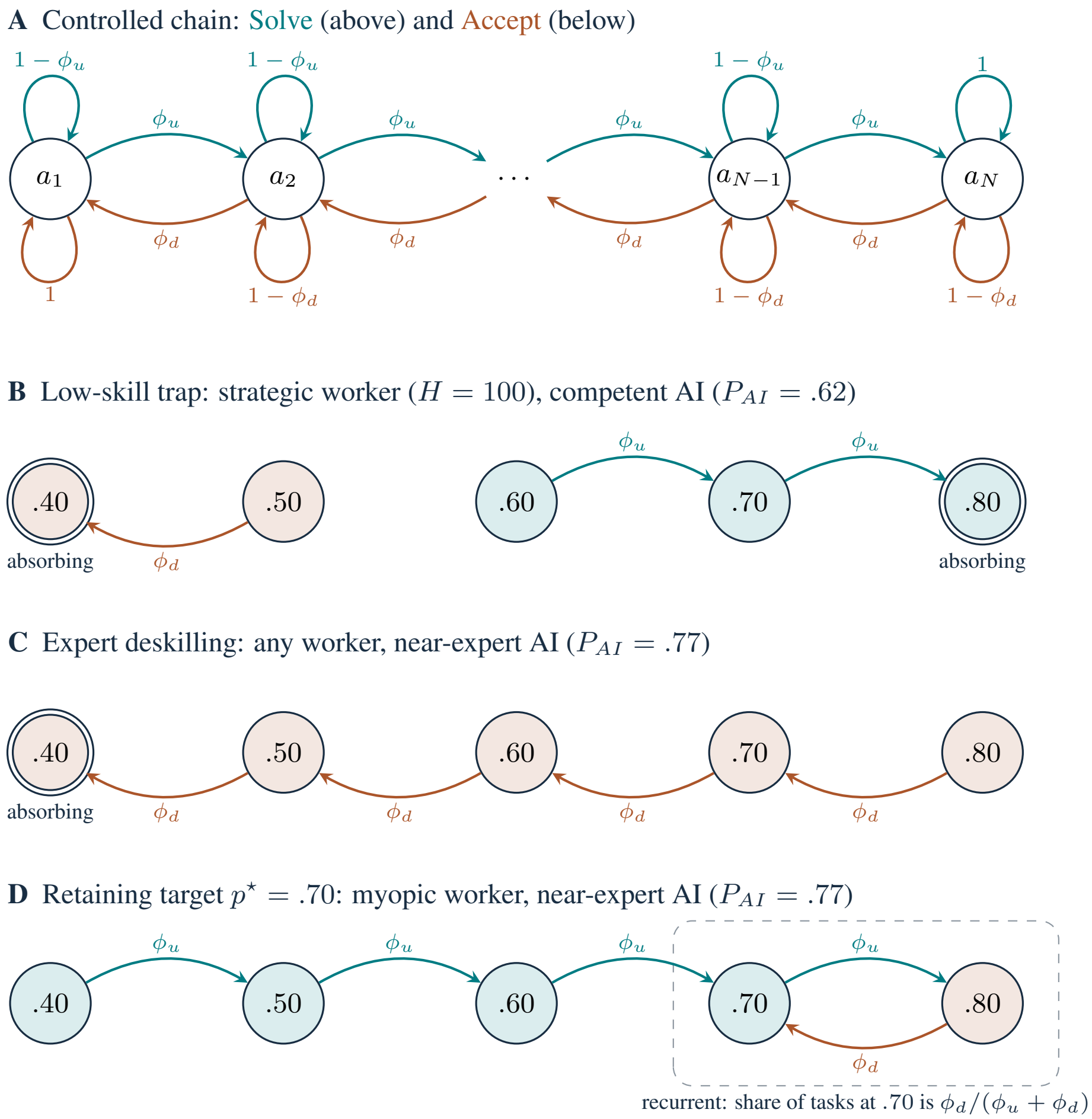
\begin{figure}[h]
\centering
\begin{tikzpicture}[
  >=stealth,
  lvl/.style={circle, draw=chainInk, line width=.6pt, minimum size=9mm, inner sep=0pt, font=\small},
  solnode/.style={lvl, fill=chainSolve!14},
  accnode/.style={lvl, fill=chainAccept!14},
  absorb/.style={double=white, double distance=1.1pt},
  sol/.style={->, chainSolve, line width=.8pt},
  acc/.style={->, chainAccept, line width=.8pt},
  lab/.style={font=\scriptsize, inner sep=1.5pt},
  ttl/.style={anchor=west, font=\small, text=chainInk},
  note/.style={font=\scriptsize, text=chainInk}]
\node[ttl] at (-.6,1.75) {\textbf{A}\enspace Controlled chain: \textcolor{chainSolve}{Solve} (above) and \textcolor{chainAccept}{Accept} (below)};
\foreach \i/\x/\t in {1/0/a_1, 2/2.6/a_2, 4/7.8/a_{N-1}, 5/10.4/a_N}
  \node[lvl] (A\i) at (\x,0) {$\t$};
\node[font=\small] (A3) at (5.2,0) {$\cdots$};
\foreach \i/\j in {1/2, 2/3, 3/4, 4/5} {
  \draw[sol] (A\i) to[bend left=30] node[lab, above] {$\phi_u$} (A\j);
  \draw[acc] (A\j) to[bend left=30] node[lab, below] {$\phi_d$} (A\i);
}
\foreach \i/\t in {1/1-\phi_u, 2/1-\phi_u, 4/1-\phi_u, 5/1}
  \draw[sol] (A\i) to[out=115, in=65, looseness=7] node[lab, above] {$\t$} (A\i);
\foreach \i/\t in {1/1, 2/1-\phi_d, 4/1-\phi_d, 5/1-\phi_d}
  \draw[acc] (A\i) to[out=-65, in=-115, looseness=7] node[lab, below] {$\t$} (A\i);
\begin{scope}[yshift=-3.6cm]
\node[ttl] at (-.6,1.2) {\textbf{B}\enspace Low-skill trap: strategic worker ($H=100$), competent AI ($\PAI=.62$)};
\foreach \i/\x/\t/\s in {1/0/.40/accnode, 2/2.6/.50/accnode, 3/5.2/.60/solnode, 4/7.8/.70/solnode, 5/10.4/.80/solnode}
  \node[\s] (B\i) at (\x,0) {$\t$};
\node[absorb, lvl, fill=chainAccept!14] at (B1) {$.40$};
\node[absorb, lvl, fill=chainSolve!14] at (B5) {$.80$};
\draw[acc] (B2) to[bend left=30] node[lab, below] {$\phi_d$} (B1);
\draw[sol] (B3) to[bend left=30] node[lab, above] {$\phi_u$} (B4);
\draw[sol] (B4) to[bend left=30] node[lab, above] {$\phi_u$} (B5);
\node[note] at (0,-.68) {absorbing};
\node[note] at (10.4,-.68) {absorbing};
\end{scope}
\begin{scope}[yshift=-6.4cm]
\node[ttl] at (-.6,1.2) {\textbf{C}\enspace Expert deskilling: any worker, near-expert AI ($\PAI=.77$)};
\foreach \i/\x/\t in {1/0/.40, 2/2.6/.50, 3/5.2/.60, 4/7.8/.70, 5/10.4/.80}
  \node[accnode] (C\i) at (\x,0) {$\t$};
\node[absorb, lvl, fill=chainAccept!14] at (C1) {$.40$};
\foreach \i/\j in {1/2, 2/3, 3/4, 4/5}
  \draw[acc] (C\j) to[bend left=30] node[lab, below] {$\phi_d$} (C\i);
\node[note] at (0,-.68) {absorbing};
\end{scope}
\begin{scope}[yshift=-9.2cm]
\draw[chainInk!50, dashed, rounded corners=6pt] (6.95,-.92) rectangle (11.25,.92);
\node[ttl] at (-.6,1.2) {\textbf{D}\enspace Retaining target $p^\star=.70$: myopic worker, near-expert AI ($\PAI=.77$)};
\foreach \i/\x/\t/\s in {1/0/.40/solnode, 2/2.6/.50/solnode, 3/5.2/.60/solnode, 4/7.8/.70/solnode, 5/10.4/.80/accnode}
  \node[\s] (D\i) at (\x,0) {$\t$};
\foreach \i/\j in {1/2, 2/3, 3/4, 4/5}
  \draw[sol] (D\i) to[bend left=30] node[lab, above] {$\phi_u$} (D\j);
\draw[acc] (D5) to[bend left=30] node[lab, below] {$\phi_d$} (D4);
\node[note] at (9.1,-1.12) {recurrent: share of tasks at $.70$ is $\phi_d/(\phi_u+\phi_d)$};
\end{scope}
\end{tikzpicture}
\caption{Skill dynamics as a Markov chain. \textbf{(A)} The controlled chain. If the worker solves, skill rises by one level with probability $\phi_u$; if the worker accepts, it falls by one level with probability $\phi_d$; otherwise it stays. At the boundaries, a step that would leave the grid is replaced by staying, so the corresponding self-loop has probability one. \textbf{(B--D)} Chains under fixed policies on the running-example grid. Fill marks the action at each level; self-loops are omitted. \textbf{(B)} A strategic worker with the competent AI accepts at $.40$ and $.50$ and solves above. No transition connects $.50$ and $.60$, so a worker starting at $.50$ can only drift to $.40$. \textbf{(C)} With the near-expert AI, every level accepts, and skill drifts to $.40$. \textbf{(D)} The target incentive for $p^\star=.70$ makes the worker solve up to the target, while the worker accepts at $.80$. Skill eventually alternates between $.70$ and $.80$; the share of tasks at $.70$ sets the long-run cost of the target (Appendix~\ref{app:incentives}). Parameters as in Section~\ref{sec:today}.}
\label{fig:markov-chain}
\end{figure}

\paragraph{Transition rule.}
Skill takes one of the $N$ levels in Eq.~\eqref{eq:skill-grid}. After each task, skill either stays on the same level or moves to an adjacent level. Which moves are possible depends on the chosen action:
\begin{align*}
\text{Solve:}&\quad p_{t+1}=\begin{cases}p^+ & \text{with probability }\phi_u,\\ p & \text{with probability }1-\phi_u,\end{cases}
&
\text{Accept:}&\quad p_{t+1}=\begin{cases}p^- & \text{with probability }\phi_d,\\ p & \text{with probability }1-\phi_d.\end{cases}
\end{align*}
At the highest skill level, a learning step is not possible, so $p^+=a_N$ and the worker stays. At the lowest level, the same holds for a decay step, $p^-=a_1$. Solving therefore never lowers skill, and accepting never raises it. Figure~\ref{fig:markov-chain}A shows these transitions.

The transition rule makes skill a finite-state Markov chain whose transitions the worker controls through the action on each task.

\emph{Finite, ordered levels.} Skill is the worker's unaided accuracy, divided into $N$ ordered levels. Knowledge tracing \citep{corbett1994} uses the simplest such division, with only two states in which a skill is either learned or not. More levels allow accuracy to rise and fall gradually. Appendix~\ref{app:robustness} shows that the results do not depend on the number or spacing of the levels.

\emph{Memoryless transitions.} The next level depends only on the current level and the current action, not on how the worker reached that level. The current level thus summarizes everything about past practice that matters for the future. The model therefore does not capture the effects of the order of past tasks, such as how practice is spaced over time.

\emph{One level per task.} Each task moves skill by at most one level, so skill changes gradually rather than in jumps. A Markov chain with this property is a birth--death chain, the structure that \citet{aouad2026} also use. In their model, learning depends on effort and decay is exogenous. In our model, both directions depend on the worker's choice on each task: only solving can raise skill, and only accepting can lower it.

\emph{Constant step probabilities.} Knowledge tracing assumes that the probability of learning is the same at every practice opportunity \citep{corbett1994}. We likewise keep $\phi_u$ and $\phi_d$ constant across levels. This keeps the model parsimonious, although the results for strategic workers do not require it: Appendix~\ref{app:closed-form} shows that the closed form extends to probabilities that differ between levels.

\emph{Holding boundaries.} At $a_1$ and $a_N$, a step that would leave the grid is replaced by staying. The self-loop at the boundary therefore has probability one under the corresponding action.

Together, these choices turn the strategic worker's problem into a finite Markov decision process \citep{puterman1994}, with $N$ states and two actions. All expectations can therefore be computed exactly from the transition matrices (Appendix~\ref{app:solution}), without simulation, and the optimal policy has a closed form (Appendix~\ref{app:closed-form}).

\paragraph{Chains under fixed policies.}
Once the worker's action at each level is fixed, only one set of arcs remains at each level, and the resulting chain shows where skill ends up in the long run (Figure~\ref{fig:markov-chain}B--D). Under a policy that accepts below some level and solves from that level upward, no transition connects the highest accepting level with the lowest solving level. The chain then splits into two parts: from accepting levels, skill can only drift down to $a_1$, and from solving levels it can only climb to $a_N$. Both $a_1$ and $a_N$ are absorbing. A worker who starts below the threshold therefore ends at $a_1$, which is the low-skill trap (panel B). If every level accepts, $a_1$ is the only absorbing level, which is expert deskilling (panel C). Under a target incentive (Eq.~\eqref{eq:incentive-target}) that leaves the worker accepting just above the target, skill eventually alternates between the target and the level above it (panel D). The long-run share of tasks spent at the target is $(1/\phi_u)/(1/\phi_u+1/\phi_d)=\phi_d/(\phi_u+\phi_d)$, which sets the long-run cost of retaining it (Appendix~\ref{app:incentives}).

\paragraph{Expected waiting time for a step.}
After each solved task, a learning step occurs with the same probability $\phi_u$, independently of earlier tasks. The number of solved tasks $K$ until the first step therefore follows a geometric distribution, $\Pr(K=k)=(1-\phi_u)^{k-1}\phi_u$ for $k=1,2,\dots$, with mean
\begin{equation}\label{eq:app-waiting}
\mathbb E[K]=\sum_{k\ge1}k\,(1-\phi_u)^{k-1}\phi_u=\frac{1}{\phi_u}.
\end{equation}
The same argument gives an average of $1/\phi_d$ accepted tasks until a decay step. To climb $r$ levels by solving, the worker waits $r$ times in a row, so by linearity of expectation the climb takes $r/\phi_u$ solved tasks on average.

\textbf{Running example.} \textit{With $\phi_u=.02$ and $\phi_d=.01$, a developer who solves rises one level after $50$ tasks on average, and one who accepts falls one level after $100$ tasks on average. Climbing from $.40$ to $.70$ takes $3\times50=150$ solved tasks on average.}

\paragraph{Expected next-task accuracy, Eq.~\eqref{eq:next-skill-gap}.}
After solving, the next-task accuracy is $p^+$ with probability $\phi_u$ and $p$ otherwise. Its expectation is therefore $(1-\phi_u)p+\phi_u p^+$. After accepting, it is $(1-\phi_d)p+\phi_d p^-$. Subtracting the second from the first and collecting terms gives
\[
\big[(1-\phi_u)p+\phi_u p^+\big]-\big[(1-\phi_d)p+\phi_d p^-\big]
=\phi_u(p^+-p)+\phi_d(p-p^-).
\]
Both terms are nonnegative because $p^+\ge p\ge p^-$. At the highest level, the first term vanishes because $p^+=p$, leaving only the avoided decay. At the lowest level, the second term vanishes.

\textbf{Running example.} \textit{At $p=.60$, solving rather than accepting raises the expected next-task accuracy by $.02\times.10+.01\times.10=.003$.}

\subsection{Expected utility of one task}\label{app:one-task}

\paragraph{Expected utilities, Eq.~\eqref{eq:payoffs-dynamic}.}
Each entry of Table~\ref{tab:utility} is weighted by the probability that the final answer is correct. Accepting submits the AI's answer, which is correct with probability $\PAI$:
\[
u_A=\PAI\cdot\gamma+(1-\PAI)\cdot0=\gamma\PAI.
\]
Solving submits the worker's own answer, which is correct with probability $p$, and costs $\lambda$ in either case:
\[
u_S(p)=p\,(\gamma-\lambda)+(1-p)(-\lambda)=\gamma p-\lambda.
\]

\paragraph{Relation to the utilities of \citet{bansal2021}.}
\citet[Table~1]{bansal2021} reward a correct answer with $1$ and penalize an incorrect one with $-\beta$, and solving also costs $\lambda$. Their four entries are $1$ and $-\beta$ for accepting and $1-\lambda$ and $-\beta-\lambda$ for solving. Adding the constant $\beta$ to every entry gives $1+\beta$ and $0$ for accepting and $1+\beta-\lambda$ and $-\lambda$ for solving, which is Table~\ref{tab:utility} with $\gamma=1+\beta$. A constant added to every outcome raises the expected utility of both actions by the same amount, so it does not change the comparison between them. Over a sequence of tasks, it adds the same constant $\beta/(1-\delta)$ to every lifetime value in Eq.~\eqref{eq:dynamic-value}, which again changes no decision.

\paragraph{Immediate payoff gap, Eq.~\eqref{eq:immediate-payoff-gap}.}
Subtracting the two expected utilities gives
\[
\Delta u(p)=u_A-u_S(p)=\gamma\PAI-(\gamma p-\lambda)=\lambda+\gamma(\PAI-p).
\]
Dividing by $\gamma>0$ does not change the sign, and $\Delta u(p)/\gamma=\lambda/\gamma+\PAI-p$. Every decision therefore depends on $\lambda$ and $\gamma$ only through their ratio $\lambda/\gamma$. The gap falls by $\gamma$ for every unit of skill, so it is largest at the lowest level.

\textbf{Running example.} \textit{With the competent AI, $\Delta u(p)=.05+.62-p$, which is $.27$, $.17$, $.07$, $-.03$, and $-.13$ at the five levels. With the near-expert AI, $\Delta u(p)=.05+.77-p$, which is $.42$, $.32$, $.22$, $.12$, and $.02$.}

\subsection{The myopic decision rule}\label{app:myopic}

A myopic worker compares only today's expected utilities and solves if $u_S(p)\ge u_A$, that is, if $\Delta u(p)\le0$. We resolve ties in favor of solving. Rearranging,
\[
\lambda+\gamma(\PAI-p)\le0
\quad\Longleftrightarrow\quad
\gamma p\ge\gamma\PAI+\lambda
\quad\Longleftrightarrow\quad
p\ge\PAI+\frac{\lambda}{\gamma},
\]
which is Eq.~\eqref{eq:myopic-rule}. Appropriate reliance solves if $p\ge\PAI$, so the myopic threshold lies higher by exactly $\lambda/\gamma$. A worker with $\PAI\le p<\PAI+\lambda/\gamma$ is more accurate than the AI but accepts. This is the effort-induced overreliance of \citet{holstein2026}.

\textbf{Running example.} \textit{The myopic threshold is $.62+.05=.67$ with the competent AI and $.77+.05=.82$ with the near-expert AI. Sam at $.80$ lies between $.77$ and $.82$ and therefore accepts, even though Sam is more accurate than the AI.}

\subsection{The discount factor as a continuation probability}\label{app:delta}

The lifetime value in Eq.~\eqref{eq:dynamic-value} weights the utility of task $t$ by $\delta^t$. Two descriptions of the worker lead to this weight. They differ in how many tasks the worker performs and in what the weight means, but not in the values or decisions they imply.

\paragraph{Discounting: the worker performs tasks indefinitely.}
In the first description, the worker performs an infinite sequence of tasks, and every task takes place with certainty. The worker, however, values a utility $t$ tasks ahead at only $\delta^t$ times the same utility today. The weight $\delta^t$ is then a preference: it states how much the worker cares about a task that will certainly happen. The weights sum to
\[
\sum_{t=0}^{\infty}\delta^t=\frac{1}{1-\delta}=H.
\]
An infinite stream of tasks with the same expected utility $u$ is therefore worth $Hu$, as much as $H$ undiscounted tasks. Under this description, $H$ is an \emph{effective} horizon. It measures how far ahead the worker looks. It is not the number of tasks the worker performs, because that number is infinite.

\paragraph{Continuation: the worker performs an expected number of tasks.}
In the second description, the worker does not discount. Every task that takes place counts fully. However, after each task another task of the same family follows only with probability $\delta$, independently of the worker's skill and choices. The task family itself continues indefinitely, but the worker's participation in it ends at a random point, for example when the worker moves to another role. The weight $\delta^t$ is then a probability: it states how likely it is that a task the worker fully cares about takes place.

Let $T$ be the number of tasks the worker performs, including the current one. Task $t$ (counting from $t=0$) takes place if and only if the $t$ continuations before it all happen, so $\Pr(T>t)=\delta^t$. The expected total utility is
\[
\mathbb E\Big[\sum_{t=0}^{T-1}U_t\Big]
=\sum_{t=0}^{\infty}\Pr(T>t)\,\mathbb E[U_t]
=\sum_{t=0}^{\infty}\delta^t\,\mathbb E[U_t].
\]
The first equality writes the random sum as a sum over all $t$, each counted only if task $t$ takes place, and uses that the continuation is independent of $U_t$. The expected number of tasks is
\[
\mathbb E[T]=\sum_{t=0}^{\infty}\Pr(T>t)=\sum_{t=0}^{\infty}\delta^t=\frac{1}{1-\delta}=H.
\]
Under this description, $H$ is the \emph{expected} horizon, the average number of tasks the worker actually performs.

\paragraph{Why the two descriptions are equivalent.}
In both descriptions, task $t$ enters the objective with weight $\delta^t$. Under discounting, a certain task receives the reduced weight $\delta^t$; under continuation, a task with full weight takes place with probability $\delta^t$. Because the worker maximizes an expectation, only the product of the two matters. Both descriptions therefore yield the objective in Eq.~\eqref{eq:dynamic-value}, the same lifetime values, and the same optimal decisions. This equivalence concerns the worker's optimization problem; it does not imply equal training-completion probabilities or equal organizational payments over an actual career. The same sum $1/(1-\delta)$ is the effective horizon under discounting and the expected number of tasks under continuation. The main text uses the continuation reading because it gives $\delta$ a concrete meaning: how many more tasks of this family the worker expects to perform. Accordingly, $H$ on the vertical axis of Figure~\ref{fig:who-solves} is the expected horizon. The myopic worker is the case $\delta=0$ in both readings. Under continuation, the current task is then certainly the last one, so $H=1$.

\paragraph{Expected versus fixed horizon.}
An expected horizon of $H$ tasks differs from a fixed horizon of exactly $H$ tasks. First, the realized number of tasks varies widely. $T$ follows a geometric distribution with $\Pr(T>m)=\delta^m$ for every integer $m\ge0$. The worker therefore performs more than $H$ tasks only with probability $\delta^{\lfloor H\rfloor}$, which is close to $e^{-1}\approx.37$ for large $H$.
Second, the expected horizon does not shrink as tasks pass. A worker who has already performed $t$ tasks still expects $\mathbb E[T-t\mid T>t]=H$ tasks, including the current one, because each continuation has the same probability $\delta$. Every task therefore looks the same to the worker, and the lifetime value depends only on the current skill level $p$, not on the task index $t$. The Bellman equation in Appendix~\ref{app:bellman} relies on this property. With a fixed horizon, the value would also depend on the number of remaining tasks. On the final task, the decision reduces to the myopic rule: solving may still be worthwhile, but future skill provides no additional benefit.

\paragraph{Finite lifetime value.}
The lifetime value is always finite. The utility of every task lies between $-\lambda$ and $\gamma$, so the sum in Eq.~\eqref{eq:dynamic-value} is bounded in absolute value by $(\gamma+\lambda)\sum_t\delta^t=(\gamma+\lambda)/(1-\delta)$.

\textbf{Running example.} \textit{With $\delta=.99$, $H=1/(1-.99)=100$. Under discounting, Anna performs code reviews indefinitely and values a bug found $100$ tasks from now at $.99^{100}\approx.37$ of a bug found today. Under continuation, Anna values both equally but expects to perform $100$ reviews, and Anna is still reviewing after $100$ tasks with probability $.99^{100}\approx.37$. Both readings give the same weight. That $H$ equals the $1/\phi_d=100$ tasks until a decay step is a coincidence of the chosen parameters.}

\subsection{The Bellman equation and the strategic decision rule}\label{app:bellman}

\paragraph{Why a Bellman equation.}
A general policy can specify actions for every possible history. In this finite discounted Markov decision process, however, an optimal deterministic stationary policy exists, so it suffices to consider the $2^N$ state-based policies used here. Bellman reasoning avoids enumerating all of them. The Bellman equation replaces this search with a comparison of two actions on the current task \citep{puterman1994}. Whatever the worker does now, the remaining tasks are best handled optimally, so their value is the lifetime value at the skill level the current action leads to. This argument requires that the remaining problem is the same at every task, which holds because the expected horizon does not shrink (Appendix~\ref{app:delta}). The Bellman equation serves two purposes in the paper. First, it turns the strategic choice into a decision rule for a single task, Eq.~\eqref{eq:future-premium}, which can be compared directly with the myopic rule and with appropriate reliance. Second, it reduces the model to $N$ equations that we solve exactly (Appendix~\ref{app:solution}).

\paragraph{Bellman equation, Eq.~\eqref{eq:dynamic-bellman}.}
The Bellman equation splits the lifetime value into the current task and everything that follows. Suppose the worker at skill $p$ accepts now and then acts optimally. The current task yields $u_A$ in expectation. The next task takes place with weight $\delta$ and starts from skill $p^-$ with probability $\phi_d$ and from $p$ otherwise. From whichever level skill reaches, the remaining problem looks exactly like the original one: the same tasks, the same AI, and the same $\delta$, only from a possibly different level. Its best achievable value is therefore $V$ at that level. Accepting now thus yields
\[
Q_A(p)=u_A+\delta\big[(1-\phi_d)V(p)+\phi_dV(p^-)\big],
\]
and by the same argument solving now yields
\[
Q_S(p)=u_S(p)+\delta\big[(1-\phi_u)V(p)+\phi_uV(p^+)\big].
\]
The worker chooses the better of the two, so $V(p)=\max\{Q_A(p),Q_S(p)\}$, which is Eq.~\eqref{eq:dynamic-bellman}. At the lowest level, $p^-=p$, so the bracket of $Q_A$ equals $V(p)$; at the highest level, $p^+=p$, and the same holds for $Q_S$. Because $V$ also appears on the right-hand side, Eq.~\eqref{eq:dynamic-bellman} is a system of $N$ equations in the $N$ unknowns $V(a_1),\dots,V(a_N)$. Appendix~\ref{app:solution} explains how we solve it.

\paragraph{Strategic decision rule, Eq.~\eqref{eq:future-premium}.}
The strategic worker solves if $Q_S(p)\ge Q_A(p)$. Writing out both sides,
\[
u_S(p)+\delta\big[(1-\phi_u)V(p)+\phi_uV(p^+)\big]
\;\ge\;
u_A+\delta\big[(1-\phi_d)V(p)+\phi_dV(p^-)\big].
\]
Moving $u_S(p)$ to the right and the discounted terms of accepting to the left, the left-hand side becomes $u_A-u_S(p)=\Delta u(p)$ and the right-hand side becomes
\[
\delta\big[(1-\phi_u)V(p)+\phi_uV(p^+)-(1-\phi_d)V(p)-\phi_dV(p^-)\big]
=\delta\big[\phi_u\big(V(p^+)-V(p)\big)+\phi_d\big(V(p)-V(p^-)\big)\big],
\]
which is Eq.~\eqref{eq:future-premium}. The step is the same as in the derivation of Eq.~\eqref{eq:next-skill-gap}, with lifetime value $V$ in place of next-task accuracy. The first term is the value of a possible learning step, and the second is the value of an avoided decay step. With $\delta=0$, the right-hand side is zero and the rule reduces to $\Delta u(p)\le0$, the myopic rule.

Both terms are nonnegative whenever higher skill is worth at least as much over the remaining tasks, $V(a_1)\le\dots\le V(a_N)$. Without incentives, Eq.~\eqref{eq:app-value} below shows this directly. The future value of independent work can therefore only make a strategic worker solve more often than a myopic one, never less.

\textbf{Running example.} \textit{With the competent AI and $\delta=.99$, the lifetime values relative to always accepting are $0$, $0$, $4.06$, $9.64$, and $13.00$ at the five levels (Appendix~\ref{app:closed-form}). At $.60$, the learning term is $.99\times.02\times(9.64-4.06)=.111$ and the retention term is $.99\times.01\times(4.06-0)=.040$. Their sum, $.151$, exceeds $\Delta u(.60)=.07$, so Anna solves at $.60$. At $.50$, the learning term is $.99\times.02\times(4.06-0)=.080$, and the retention term is zero. This is less than $\Delta u(.50)=.17$, so Anna accepts at $.50$.}

\subsection{Solving the model}\label{app:solution}

\paragraph{Transition matrices.}
For a fixed action at every level, the transition rule of Appendix~\ref{app:skill-dynamics} is a matrix whose entry in row $n$ and column $m$ is the probability of moving from $a_n$ to $a_m$. Solving at every level gives $T_S$, with $1-\phi_u$ on the diagonal and $\phi_u$ just to the right of it. Accepting at every level gives $T_A$, with $1-\phi_d$ on the diagonal and $\phi_d$ just to the left of it. In the first row of $T_A$ and the last row of $T_S$, the step probability is added to the diagonal, because the worker stays at the boundary. A \emph{policy} $\pi$ picks one action per level. Its transition matrix $T_\pi$ takes row $n$ from $T_S$ or $T_A$ according to the action at $a_n$, and $u_\pi$ is the vector of the corresponding expected utilities $u_S(a_n)$ or $u_A$.

\paragraph{Value of a fixed policy.}
For a fixed policy, the maximum in Eq.~\eqref{eq:dynamic-bellman} disappears, and the equation becomes linear, $V_\pi=u_\pi+\delta T_\pi V_\pi$. Its solution is
\begin{equation}\label{eq:app-policy-value}
V_\pi=(I-\delta T_\pi)^{-1}u_\pi,
\end{equation}
where $I$ is the identity matrix. The inverse exists because $\delta<1$ and the rows of $T_\pi$ sum to one. At a single level, if the worker solves at $a_n<a_N$, then $V_\pi(a_n)=u_S(a_n)+\delta[(1-\phi_u)V_\pi(a_n)+\phi_uV_\pi(a_{n+1})]$, so
\[
V_\pi(a_n)=\frac{u_S(a_n)+\delta\phi_u V_\pi(a_{n+1})}{1-\delta+\delta\phi_u}.
\]

\paragraph{Policy iteration.}
We find the optimal policy by policy iteration \citep{puterman1994}. We start from the myopic policy, compute its value with Eq.~\eqref{eq:app-policy-value}, and then switch the action at every level at which the other action has a strictly higher value $Q_A$ or $Q_S$. We repeat these two steps until no level switches. Each round weakly improves the value at every level. With only $2^N$ possible policies, the procedure stops after finitely many rounds (in practice, after two or three), and the final policy satisfies Eq.~\eqref{eq:dynamic-bellman}. Ties are resolved in favor of solving. Value iteration, which applies the right-hand side of Eq.~\eqref{eq:dynamic-bellman} repeatedly, reaches the same solution but approaches it only at rate $\delta$ per round, which is slow for $\delta=.99$. Policy iteration is exact up to floating-point precision.

\paragraph{Expected trajectories.}
The figures show exact expectations over time rather than averages of simulated workers. Let $\mu_t$ be the row vector of probabilities that the worker is at each level before task $t$, with $\mu_0$ placing all mass on the starting level. Under a fixed policy, $\mu_{t+1}=\mu_tT_\pi$. Expected unaided accuracy before task $t$ is $\sum_n\mu_t(n)\,a_n$. Expected team accuracy, the accuracy of the submitted answer, replaces $a_n$ by $\PAI$ at levels where the worker accepts. Expected utility and expected incentive payments are computed in the same way from the per-level values.
Under the continuation interpretation, these trajectories are conditional on the worker still participating at task $t$. Undiscounted organizational costs are reported separately as benchmarks for continued participation over an unlimited task sequence.

\subsection{A closed form for the strategic worker}\label{app:closed-form}

Without incentives, the strategic worker's choice has a simple closed form. It explains the shape of Figure~\ref{fig:who-solves} and the running-example values above.

\paragraph{Expected future accuracy.}
Consider a worker who starts at level $a_n$ and solves on every task. Skill then only rises. Define the worker's \emph{expected future accuracy} as the weighted average of unaided accuracy over all future tasks, with task $t$ weighted by $(1-\delta)\delta^t$:
\[
\hat a_n=(1-\delta)\sum_{t=0}^{\infty}\delta^t\,\mathbb E\big[p_t\,\big|\,p_0=a_n,\ \text{solve on every task}\big].
\]
The weights sum to one, so $\hat a_n$ lies between $a_n$ and $a_N$, and $\hat a_N=a_N$. Conditioning on the first task gives $\hat a_n=(1-\delta)a_n+\delta[(1-\phi_u)\hat a_n+\phi_u\hat a_{n+1}]$, and solving for $\hat a_n$,
\begin{equation}\label{eq:app-ahat}
\hat a_n=(1-\omega)\,a_n+\omega\,\hat a_{n+1},
\qquad
\omega=\frac{\delta\phi_u}{1-\delta+\delta\phi_u}.
\end{equation}
The weight $\omega$ is the expected discount until the next learning step, $\mathbb E[\delta^K]$ for the geometric waiting time $K$ of Eq.~\eqref{eq:app-waiting}. Starting from $\hat a_N=a_N$, Eq.~\eqref{eq:app-ahat} gives all other levels by working downward. The lifetime utility of solving on every task is $\sum_t\delta^t\,\mathbb E[\gamma p_t-\lambda]=(\gamma\hat a_n-\lambda)/(1-\delta)$, and that of accepting on every task is $\gamma\PAI/(1-\delta)$.

\paragraph{Solving rule.}
The strategic worker's choice turns out to be a comparison between exactly these two plans. Intuitively, the levels at which the worker solves are all levels from some threshold upward. A worker who solves stays among these levels, because solving never lowers skill, and therefore solves on every future task. A worker who accepts stays among the accepting levels, because accepting never raises skill, and therefore accepts on every future task. Comparing the two plans gives the strategic decision rule in closed form,
\begin{equation}\label{eq:app-strategic-rule}
\text{solve at }a_n
\quad\Longleftrightarrow\quad
\hat a_n\;\ge\;\PAI+\frac{\lambda}{\gamma},
\end{equation}
and the lifetime value relative to always accepting,
\begin{equation}\label{eq:app-value}
V(a_n)-\frac{\gamma\PAI}{1-\delta}
=\frac{\gamma\,\big[\hat a_n-\PAI-\lambda/\gamma\big]^+}{1-\delta}.
\end{equation}
The strategic worker thus applies the myopic rule~\eqref{eq:myopic-rule} with expected future accuracy $\hat a_n$ in place of current skill $p$. With $\delta=0$, $\hat a_n=a_n$ and the rule is the myopic one. As $\delta$ approaches one, $\hat a_n$ approaches $a_N$. Eq.~\eqref{eq:app-value} increases in $n$, so higher skill is never worth less, as used in Appendix~\ref{app:bellman}.

\paragraph{Why the closed form is exact.}
We show that the values in Eq.~\eqref{eq:app-value} satisfy the Bellman equation~\eqref{eq:dynamic-bellman} at every level, and that the rule~\eqref{eq:app-strategic-rule} attains its maximum. This suffices because the Bellman equation has only one solution. Applying its right-hand side to two candidate value functions shrinks the largest difference between them by at least the factor $\delta<1$, so two different candidates cannot both remain unchanged \citep{puterman1994}. A policy that attains the maximum at every level is then optimal.

We write $A=\gamma\PAI/(1-\delta)$ for the value of always accepting, $W_n=(\gamma\hat a_n-\lambda)/(1-\delta)$ for the value of always solving from $a_n$, and $V_n=\max\{A,W_n\}$ for the candidate value. The argument uses three facts.
\begin{enumerate}
\item $\hat a_n$ increases in $n$. By Eq.~\eqref{eq:app-ahat}, $\hat a_n$ is a weighted average of $a_n$ and $\hat a_{n+1}$, and $a_n<a_{n+1}\le\hat a_{n+1}$. The levels at which $W_n\ge A$, where Eq.~\eqref{eq:app-strategic-rule} solves, are therefore all levels from some threshold upward, and $V_n$ is nondecreasing in $n$.
\item $W$ satisfies the solving branch, $W_n=u_S(a_n)+\delta[(1-\phi_u)W_n+\phi_uW_{n+1}]$, with $W_{N+1}=W_N$. This follows from the recursion that defines $\hat a_n$ after multiplying it by $\gamma$, subtracting $\lambda=(1-\delta)\lambda+\delta\lambda$ on both sides, and dividing by $1-\delta$.
\item $A$ satisfies the accepting branch with itself, $A=u_A+\delta A$.
\end{enumerate}

\emph{Levels at which the rule solves} ($W_n\ge A$). By fact~1, the level above also solves, so $V_n=W_n$ and $V_{n+1}=W_{n+1}$, and by fact~2 the solving branch equals $V_n$. The accepting branch is at most $V_n$:
\[
u_A+\delta\big[(1-\phi_d)V_n+\phi_dV_{n-1}\big]\;\le\;u_A+\delta V_n\;\le\;(1-\delta)V_n+\delta V_n=V_n.
\]
The first step uses that $V$ is nondecreasing, the second that $u_A=(1-\delta)A\le(1-\delta)V_n$. Solving therefore attains the maximum.

\emph{Levels at which the rule accepts} ($W_n<A$). By fact~1, every lower level also accepts, so $V_n=V_{n-1}=A$, and by fact~3 the accepting branch equals $V_n$. For the solving branch, note first that $u_S(a_n)<u_A$, because $a_n\le\hat a_n<\PAI+\lambda/\gamma$. If the level above also accepts, $V_{n+1}=A$, and the solving branch is $u_S(a_n)+\delta A<u_A+\delta A=A$. If the level above solves, $V_{n+1}=W_{n+1}$, and subtracting the equation of fact~2 from the solving branch gives
\[
u_S(a_n)+\delta\big[(1-\phi_u)A+\phi_uW_{n+1}\big]
=W_n+\delta(1-\phi_u)(A-W_n)
<A,
\]
because $W_n<A$ and $\delta(1-\phi_u)<1$. Accepting therefore attains the maximum.

Hence the values of Eq.~\eqref{eq:app-value} solve the Bellman equation, and the rule~\eqref{eq:app-strategic-rule} is optimal. We also confirmed Eqs.~\eqref{eq:app-strategic-rule} and~\eqref{eq:app-value} against the exact solution of Eq.~\eqref{eq:dynamic-bellman} for the running example and for every configuration of the robustness analysis in Appendix~\ref{app:robustness}.

\paragraph{Level-dependent step probabilities.}
The argument does not use that $\phi_u$ and $\phi_d$ are the same at every level. If the learning probability at $a_n$ is $\phi_{u,n}$ and the decay probability is $\phi_{d,n}$, Eq.~\eqref{eq:app-ahat} holds with the level-specific weight $\omega_n=\delta\phi_{u,n}/(1-\delta+\delta\phi_{u,n})$ in place of $\omega$, and facts~1--3 and both cases go through unchanged. The strategic rule~\eqref{eq:app-strategic-rule} and the values~\eqref{eq:app-value} therefore also hold when learning slows down or speeds up as workers progress, and decay still does not affect the choice.

The decay probability $\phi_d$ enters neither fact and therefore does not appear in Eq.~\eqref{eq:app-strategic-rule}. The retention term in Eq.~\eqref{eq:future-premium} is nonnegative. It is strictly positive when $\delta>0$ and the current level has greater value than its lower neighbor. It is zero at the lowest grid level and can also vanish at a threshold tie. It does not, however, move the boundary between solving and accepting. Decay determines how fast skill falls at the levels where the worker accepts, not which levels these are.

\textbf{Running example.} \textit{With $\delta=.99$ and $\phi_u=.02$, $\omega=.664$, and Eq.~\eqref{eq:app-ahat} gives $\hat a_n=.559$, $.640$, $.711$, $.766$, and $.80$ at the five levels. A novice at $.40$ thus decides as if their accuracy were $.559$. With the competent AI, the threshold is $.67$, so the strategic worker solves at $.60$, $.70$, and $.80$ and accepts at $.40$ and $.50$. Eq.~\eqref{eq:app-value} gives the values $(.711-.67)/.01=4.06$, $9.64$, and $13.00$ at the three solving levels. With the near-expert AI, the threshold is $.82$, above every $\hat a_n$, so the strategic worker accepts everywhere.}

\paragraph{Policy regions in Figure~\ref{fig:who-solves}.}
For a fixed starting level, Eq.~\eqref{eq:app-strategic-rule} says that the strategic worker solves if and only if $\PAI\le\hat a_n-\lambda/\gamma$, where $\hat a_n$ depends on the horizon through $\delta=1-1/H$. This gives the boundary between the shaded regions. For the novice at $.50$, the boundary lies at $\PAI=.45$ for $H=1$, at $\PAI=.59$ for $H=100$, and approaches $a_N-\lambda/\gamma=.75$ as $H$ grows. With the competent AI ($.62>.59$), Anna therefore still accepts at $H=100$. For the expert at $a_N=.80$, $\hat a_N=a_N$ at every horizon, so the boundary is the vertical line $\PAI=.75$.

\subsection{Low-skill trap and expert deskilling}\label{app:obstacles}

\paragraph{Low-skill trap.}
Suppose the worker accepts at the two lowest levels $a_1$ and $a_2$. Starting from either level, skill can only stay the same or fall, so the worker never leaves these two levels, accepts on every task, and earns $u_A$ per task. Both levels therefore have the same value, $V(a_1)=V(a_2)=\gamma\PAI/(1-\delta)$. At $a_1$, the learning term in Eq.~\eqref{eq:future-premium} is $\delta\phi_u[V(a_2)-V(a_1)]=0$, and the retention term is zero because $a_1^-=a_1$. A single learning step is worth nothing to a worker who would still accept one level higher. By Eq.~\eqref{eq:app-strategic-rule}, this situation arises exactly when $\hat a_2<\PAI+\lambda/\gamma$.

\textbf{Running example.} \textit{With the competent AI, $\hat a_2=.640<.67$, so Anna accepts at $.40$ and $.50$, and $V(.40)=V(.50)$. Starting at $.50$, expected skill after $t$ tasks is $.40+.10\times(.99)^t$, which falls toward $.40$, while output accuracy stays at $.62$, because every submitted answer is the AI's.}

\paragraph{Expert deskilling.}
Suppose $a_N<\PAI+\lambda/\gamma$. Then $\Delta u(a_N)>0$, and because $\Delta u$ decreases in skill, $\Delta u(p)>0$ at every level: at every level and on every task, accepting pays strictly more today than solving. No sequence of choices can therefore earn more than $u_A$ on any task, and always accepting earns exactly $u_A$ on every task. Hence
\[
V(p)=\sum_{t=0}^{\infty}\delta^t\gamma\PAI=\frac{\gamma\PAI}{1-\delta}\qquad\text{for every }p\in\mathcal P.
\]
Since $V$ is the same at every level, both differences in Eq.~\eqref{eq:future-premium} are zero, and the worker accepts everywhere for any $\delta$.

While the worker always accepts, let $B_t$ count potential decay events over $t$ tasks:
\[
B_t\sim\operatorname{Binomial}(t,\phi_d).
\]
Starting at $a_N$, the actual number of downward moves is capped by the lower boundary:
\[
D_t=\min\{B_t,N-1\},\qquad p_t=a_{N-D_t}.
\]
For equally spaced levels with distance $s$,
\[
\mathbb E[p_t]
=a_N-s\,\mathbb E[\min\{B_t,N-1\}].
\]
Before the lower boundary can affect the process, expected skill therefore falls by $\phi_d s$ per task. For $N\ge2$, the probability of remaining at the highest level is $(1-\phi_d)^t$. For $0<\phi_d<1$, its half-life is $\ln2/[-\ln(1-\phi_d)]$, approximately $.69/\phi_d$ for small $\phi_d$. If $\phi_d=1$, the worker leaves the highest level after the first task.

\textbf{Running example.} \textit{With the near-expert AI, $.80<.77+.05=.82$, so Sam accepts at every level. Sam is still at $.80$ with probability one half after $69$ tasks. Expected skill falls by $.01\times.10=.001$ per task at first and reaches $.70$ after about $100$ tasks, while output accuracy stays at $.77$.}

\subsection{Incentives}\label{app:incentives}

\paragraph{Incentive compatibility.}
Under the incentive, a correct independent answer earns $\gamma+\theta(p)$ instead of $\gamma$. Taking expectations with $\Pr(Y_t=1)=p$ gives $(\gamma+\theta(p))p-\lambda=u_S(p)+p\,\theta(p)$ for solving, while accepting still gives $u_A$. The myopic worker solves if
\[
u_S(p)+p\,\theta(p)\ge u_A
\quad\Longleftrightarrow\quad
p\,\theta(p)\ge u_A-u_S(p)=\Delta u(p),
\]
which is Eq.~\eqref{eq:incentive-ic}. The product $p\,\theta(p)$ is the incentive the worker expects per solved task, and it is also what the organization expects to pay per solved task.

\paragraph{Effort incentive, Eq.~\eqref{eq:effort-incentive-payoff}.}
With $\theta^{\mathrm{eff}}(p)=\lambda/p$, the expected payment per solved task is $p\cdot\lambda/p=\lambda$, so $u_S^{\mathrm{eff}}(p)=\gamma p-\lambda+\lambda=\gamma p$. The myopic worker solves if $\gamma p\ge\gamma\PAI$, that is, if $p\ge\PAI$, which is appropriate reliance. The effort incentive acts as if solving cost no effort. Repeating the argument for expert deskilling in Appendix~\ref{app:obstacles} with $\lambda=0$ shows that the expert accepts at every level once $a_N<\PAI$, so the effort incentive raises the AI accuracy at which experts start to deskill from $a_N-\lambda/\gamma$ to $a_N$. For a strategic worker, Eq.~\eqref{eq:app-strategic-rule} with $\lambda=0$ gives solving if and only if $\hat a_n\ge\PAI$.

\textbf{Running example.} \textit{Under the effort incentive, Sam at $.80$ solves with the near-expert AI ($.80\ge.77$) and receives $.05$ per task in expectation. With the above-expert AI, $.80<.82$, so Sam accepts at every level. Anna's levels $.40$, $.50$, and $.60$ all lie below $.62$, so a myopic Anna accepts at each of them.}

\paragraph{Learning incentive, Eqs.~\eqref{eq:minimum-current-incentive} and~\eqref{eq:incentive-split}.}
Eq.~\eqref{eq:incentive-ic} requires $\theta(p)\ge\Delta u(p)/p$. If $\Delta u(p)\le0$, the worker solves without any incentive, and the smallest nonnegative incentive is zero. Otherwise it is $\Delta u(p)/p$. Both cases together give $\theta^{\min}(p)=[\Delta u(p)]^+/p$. Where $\Delta u(p)\ge0$, splitting the numerator $\Delta u(p)=\lambda+\gamma(\PAI-p)$ and dividing each part by $p$ gives Eq.~\eqref{eq:incentive-split}. The expected payment per solved task is $p\,\theta^{\min}(p)=[\Delta u(p)]^+$, the positive part of the immediate payoff gap. Under the learning incentive, the expected utility of solving becomes $u_S(p)+[\Delta u(p)]^+=\max\{u_A,u_S(p)\}$, so it matches accepting where accepting would pay more and leaves the worker's payoff unchanged otherwise.

\textbf{Running example.} \textit{Anna at $.60$ with the competent AI receives $\theta^{\min}=.07/.60=.117$ per correct independent answer, or $.07$ per task in expectation, of which $.05$ is for effort and $.02$ for the accuracy gap. Sam at $.80$ with the near-expert AI receives $.02$ per task in expectation, and $.07$ with the above-expert AI.}

\paragraph{From myopic to strategic workers.}
Under an incentive, the solving branch of the Bellman equation~\eqref{eq:dynamic-bellman} gains the expected payment $p\,\theta(p)$, and the strategic worker solves if $p\,\theta(p)+F(p)\ge\Delta u(p)$, where $F(p)$ denotes the right-hand side of Eq.~\eqref{eq:future-premium} computed with the lifetime values under the incentive. Where the incentive moves a myopic worker, $p\,\theta(p)\ge\Delta u(p)$ already holds, so the strategic worker also solves as long as $F(p)\ge0$, that is, as long as higher skill is worth at least as much under the incentive. This holds for the incentives in Section~\ref{sec:dynamic-incentives} in the running example and in every configuration of the robustness analysis (Appendix~\ref{app:robustness}).

\paragraph{Target incentive, Eq.~\eqref{eq:incentive-target}.}
The target incentive is the learning incentive paid only at levels $p\le p^\star$. At and below the target, the myopic worker therefore solves. Above it, the worker faces the unpaid choice and solves only if $\Delta u(p)\le0$.

\paragraph{Cost of acquiring a target.}
Organizational costs in Section~\ref{sec:dynamic-incentives} are expected payments to a myopic worker, summed over an unlimited sequence of tasks without discounting. These are not expected payments over a geometrically terminated career. Suppose the worker starts at $a_i$ with $i\le j<N$, the target is $p^\star=a_j$, and solving pays without an incentive one level above it, $\Delta u(a_{j+1})\le0$. Because $\Delta u$ decreases in skill, solving then pays on its own at every level above the target as well. The worker solves at every level from $a_i$ upward, so skill never falls. At each supported level $a_n$, the worker spends on average $1/\phi_u$ tasks (Eq.~\eqref{eq:app-waiting}) and receives $[\Delta u(a_n)]^+$ per task in expectation. Whether a learning step happens does not depend on whether the answer was correct, so the expected payment while at $a_n$ is the product of the two. Summing over the supported levels,
\begin{equation}\label{eq:app-acquisition-cost}
\text{expected total payment}=\frac{1}{\phi_u}\sum_{n=i}^{j}\big[\Delta u(a_n)\big]^+ .
\end{equation}
After the worker reaches $a_{j+1}$, payments stop permanently. If instead $\Delta u(a_{j+1})>0$, the worker accepts just above the target, eventually decays back to it, and the incentive is paid again on every return, so payments recur indefinitely.

For the top target $j=N$, payments end if $\Delta u(a_N)\le0$. In this case, the expected total payment is the sum of the transient-state costs,
\[
\frac{1}{\phi_u}
\sum_{n=i}^{N-1}[\Delta u(a_n)]^+,
\]
with an empty sum equal to zero. If $\Delta u(a_N)>0$, the top state requires a positive expected payment on every task. A worker already in an unpaid invariant solving region requires no payments, irrespective of a lower target.

\textbf{Running example.} \textit{With the competent AI, starting at $.40$ with target $.60$, solving pays on its own at $.70$ ($\Delta u(.70)=-.03$), and Eq.~\eqref{eq:app-acquisition-cost} gives $50\times(.27+.17+.07)=25.50$. Starting at $.50$ with the learning incentive at every level, as in Figure~\ref{fig:incentive-mechanisms}A, it gives $50\times(.17+.07)=12.00$. With target $.50$, Anna accepts at $.60$ and payments recur: each return to $.50$ costs $50\times.17=8.50$ and is followed by $100$ unpaid tasks at $.60$ on average, a long-run rate of $8.50/150\approx.057$ per task. The same counting gives the additional errors of practice in Section~\ref{sec:ethics}. While practicing at $.40$, $.50$, and $.60$ instead of accepting, Anna submits $50\times[(.62-.40)+(.62-.50)+(.62-.60)]=18$ more incorrect answers in expectation than the AI would.}

\paragraph{Cost of retaining a target.}
At the target, the worker solves, so skill cannot fall below it. Above the target, accepting lowers skill by at most one level per task, so skill must pass through the target before it could fall further, and there the worker solves again. If the target is the highest level, $p^\star=a_N$, the worker solves on every task and the organization pays $[\Delta u(a_N)]^+$ per task for as long as the skill is to be retained.

If the target is one level lower, $p^\star=a_{N-1}$, and the worker accepts at $a_N$ ($\Delta u(a_N)>0$), skill alternates between the two levels. Each cycle consists of on average $1/\phi_d$ unpaid tasks at $a_N$ and $1/\phi_u$ supported tasks at $a_{N-1}$, each paying $\Delta u(a_{N-1})$ in expectation. The long-run payment per task is the expected payment per cycle divided by the expected length of a cycle,
\[
\frac{\Delta u(a_{N-1})/\phi_u}{1/\phi_u+1/\phi_d}=\Delta u(a_{N-1})\,\frac{\phi_d}{\phi_u+\phi_d}.
\]
The higher target is therefore cheaper in the long run if $\Delta u(a_N)(\phi_u+\phi_d)<\Delta u(a_{N-1})\,\phi_d$. With equally spaced levels at distance $s$, $\Delta u(a_{N-1})=\Delta u(a_N)+\gamma s$, and the condition simplifies to $\Delta u(a_N)\,\phi_u<\gamma s\,\phi_d$. This condition holds when the expert's remaining payoff gap is small relative to the larger gap one level down.

\textbf{Running example.} \textit{With the near-expert AI, target $.80$ costs $\Delta u(.80)=.02$ per task. Target $.70$ costs $\Delta u(.70)=.12$ per supported task, and a cycle consists of $100$ unpaid and $50$ supported tasks on average, so the long-run payment is $.12\times50/150=.04$ per task. The condition reads $.02\times.02=.0004<.10\times.01=.001$.}

\subsection{Strategic return of independent work}\label{app:strategic-return}

\begin{figure}[h]
    \centering
    \includegraphics[width=0.6\linewidth]{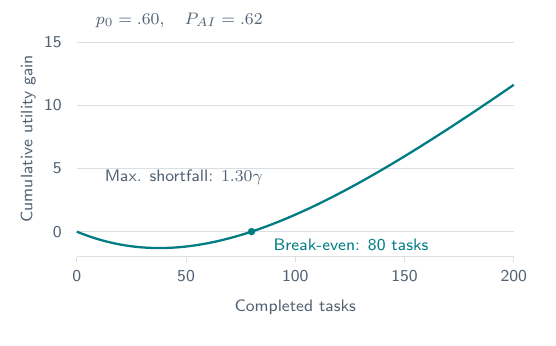}
    \caption{Strategic independent work pays off despite an initial
    cost. Starting at $p_0=.60$ with the competent AI
    ($\PAI=.62$), the curve shows expected cumulative undiscounted
    utility under the strategic policy minus that under the
    myopic policy. The largest shortfall is $1.30\gamma$ after
    38 tasks, and the gain first becomes nonnegative after
    80 completed tasks. The discounted lifetime gain is
    $4.06\gamma$. Parameters as in Section~\ref{sec:today},
    with $\delta=.99$.}
    \label{fig:strategic-return}
\end{figure}

Figure~\ref{fig:strategic-return} traces the running example of Section~\ref{sec:strategic} over time. Starting at $p_0=.60$ with the competent AI, the strategic worker solves and initially earns less than a myopic worker, who accepts. For both policies we compute the expected utility of each task as in Appendix~\ref{app:solution} and plot the cumulative difference without discounting. The strategic worker gives up $\Delta u(.60)=.07$ per task at first. As skill rises, the shortfall per task shrinks and turns into a gain. The cumulative shortfall is largest after 38 tasks and is offset after 80 tasks. Thereafter the gain grows by up to $.80-.05-.62=.13$ per task, the advantage of an expert who solves over a worker who accepts. Weighting task $t$ by $\delta^t$ instead gives the discounted lifetime gain of $4.06$, which equals the closed-form value $(.711-.67)/.01$ of Eq.~\eqref{eq:app-value}, because a myopic worker at $.60$ accepts on every task.

\section{Robustness analysis}\label{app:robustness}

Sections~\ref{sec:long-horizon} and~\ref{sec:dynamic-incentives} illustrate the results with one set of parameters. This appendix shows that the qualitative results do not depend on that choice. We vary every parameter of the model over a wide range, solve the model exactly for every combination, and check each qualitative result (Appendix~\ref{app:rob-checks}). Because the decisions depend on the parameters only through a few combinations, the results can also be summarized in two figures (Appendix~\ref{app:rob-regions}). Finally, we vary the running example one parameter at a time to show which of its specific outcomes depend on the chosen values (Appendix~\ref{app:rob-oat}).

\subsection{Design}\label{app:rob-design}

Table~\ref{tab:rob-params} lists the parameter values. We combine all of them in a full factorial design. The reward is fixed at $\gamma=1$ because decisions depend on $\lambda$ and $\gamma$ only through $\lambda/\gamma$ (Appendix~\ref{app:one-task}). The eight skill grids vary the number of levels, the range of accuracies, and the spacing between levels. With a constant step probability, unequal spacing makes the accuracy gained per learning step change with skill, which stands in for learning that slows down or speeds up as workers progress. The learning and decay probabilities are stated per step of $.10$, as in the running example. On grids with a different average step, we multiply the probabilities by $.10$ divided by the mean step size. This preserves interior expected accuracy drift on uniform grids. On unequal grids, local drift still depends on the size of the adjacent step. A grid with steps of $.05$ thus uses $2\phi_u$ and $2\phi_d$. Combinations in which a scaled probability would exceed one are dropped.

\begin{table}[h]
\caption{Parameter values in the full factorial design. Running-example values in bold.}
\label{tab:rob-params}
\small
\begin{tabular}{@{}lL{8.2cm}r@{}}
\toprule
Parameter & Values & Count\\
\midrule
Effort cost $\lambda/\gamma$ & $.01$, $.025$, $\mathbf{.05}$, $.10$, $.20$ & 5\\
Learning probability $\phi_u$ (per step of $.10$) & $.005$, $.01$, $\mathbf{.02}$, $.05$, $.10$, $.25$ & 6\\
Decay probability $\phi_d$ (per step of $.10$) & $.0025$, $.005$, $\mathbf{.01}$, $.02$, $.05$, $.10$ & 6\\
Horizon $H=1/(1-\delta)$ & $2$, $5$, $10$, $20$, $50$, $\mathbf{100}$, $200$, $500$, $1000$ & 9\\
AI accuracy $\PAI$ & $.30$, $.31$, \dots, $.96$ & 67\\
Skill grid & \textbf{5 levels on $[.40,.80]$}; 3, 9, or 17 equally spaced levels on $[.40,.80]$; 7 levels on $[.30,.90]$; 8 levels on $[.60,.95]$; diminishing steps $\{.40,.50,.59,.67,.74,.80\}$; increasing steps $\{.40,.44,.50,.58,.68,.80\}$ & 8\\
\bottomrule
\end{tabular}
\end{table}

This gives 1{,}440 combinations of grid, effort cost, and step probabilities, each solved for 67 AI accuracies and 9 horizons, or 868{,}320 configurations in total. For each configuration, we compute the myopic policy and the strategic policy by exact policy iteration (Appendix~\ref{app:solution}), without incentives, with the effort incentive, with the learning incentive at every level, and with a target incentive for every possible target. Incentive costs are numerical expectations computed from the transition matrices for myopic and strategic workers. For the sampled policies, long-run payment rates are computed by repeated squaring of the transition matrix, and total payments use expected transient-state visit counts. The policy comparison tolerance is $10^{-9}$; the closed-form policy check excludes levels within $10^{-8}$ of a threshold. For periodic chains permitted at the boundary of the model's parameter domain, long-run payment rates are interpreted as time averages rather than pointwise limits. Ties are resolved in favor of solving, as in the main text.

\subsection{Which results hold in every configuration}\label{app:rob-checks}

Table~\ref{tab:rob-checks} lists the checks. Every check holds in every configuration.

\begin{table}[h]
\caption{Qualitative results checked in the full factorial design. Each row counts the configurations in which the condition applies and those in which it fails. Rows~1--7 concern workers without incentives, rows~8--14 the incentives of Section~\ref{sec:dynamic-incentives}. Rows~11 and~13 count every target in every configuration; rows~8, 12, and~14 do not depend on the horizon; row~7 counts combinations of the other parameters, each compared across the six decay probabilities.}
\label{tab:rob-checks}
\small
\begin{tabular}{@{}rL{10.6cm}rr@{}}
\toprule
& Result & Checked & Failed\\
\midrule
1 & Strategic policy is a threshold in skill (solve at $a_n\Rightarrow$ solve at all $a_m>a_n$) & 868{,}320 & 0\\
2 & Strategic policy equals the closed form, Eq.~\eqref{eq:app-strategic-rule}$^{a}$ & 868{,}320 & 0\\
3 & Lifetime values equal the closed form, Eq.~\eqref{eq:app-value} & 868{,}320 & 0\\
4 & $V$ is nondecreasing in skill; the strategic worker solves wherever the myopic one does & 868{,}320 & 0\\
5 & Low-skill trap: if the worker accepts at $a_1$ and $a_2$, then $V(a_1)=V(a_2)$ & 520{,}524 & 0\\
6 & Expert deskilling: all levels accept and $V$ is constant iff $a_N<\PAI+\lambda/\gamma$ & 868{,}320 & 0\\
7 & Strategic policy does not depend on $\phi_d$ & 144{,}720 & 0\\
\midrule
8 & Effort incentive: myopic worker solves iff $p\geq\PAI$ (appropriate reliance) & 96{,}480 & 0\\
9 & Effort incentive: strategic worker solves iff $\hat a_n\geq\PAI$; all accept iff $a_N<\PAI$ & 868{,}320 & 0\\
10 & Learning incentive $\theta^{\min}$ at every level moves the strategic worker to solve everywhere & 868{,}320 & 0\\
11 & Target incentive moves the strategic worker to solve at every level $\leq p^\star$ & 6{,}620{,}940 & 0\\
12 & Myopic target policy: payment-ending conditions and acquisition costs or recurrent payment rates (App.~\ref{app:incentives}) & 735{,}660 & 0\\
13 & Strategic workers never cost more in the long run than myopic workers under the same target & 6{,}620{,}940 & 0\\
14 & Retention: long-run rates of targets $a_N$ and $a_{N-1}$ and the cost condition of App.~\ref{app:incentives} & 29{,}772 & 0\\
\bottomrule
\end{tabular}
\\[2pt]\parbox{\linewidth}{\footnotesize $^{a}$\,Up to numerical ties. In 6 configurations (17 levels, $\lambda/\gamma=.025$, $\phi_u=.10$, $H=2$, $\PAI=.46$, any $\phi_d$), $\hat a_n$ falls short of $\PAI+\lambda/\gamma$ by less than $10^{-9}$ at one level, and the two computations break this near-tie differently.}
\end{table}

\paragraph{Strategic decisions.}
Rows~1--4 confirm the closed form of Appendix~\ref{app:closed-form} across all configurations: the strategic worker solves exactly at the levels at which expected future accuracy $\hat a_n$ is at least $\PAI+\lambda/\gamma$, and lifetime values follow Eq.~\eqref{eq:app-value}. This closed form reduces the parameter space. Decay $\phi_d$ does not enter it (row~7), and the learning probability and the horizon enter only through the weight $\omega$ of Eq.~\eqref{eq:app-ahat}. With $\delta=1-1/H$,
\[
\omega=\frac{\delta\phi_u}{1-\delta+\delta\phi_u}=\frac{\kappa}{1+\kappa},
\qquad
\kappa=\phi_u(H-1).
\]
Since $H-1$ is the expected number of tasks after the current one, $\kappa$ is the expected number of learning steps a worker who always solves could take over the remaining tasks, ignoring the top level. A slow learner with a long horizon therefore decides exactly like a fast learner with a short one if both expect the same number of learning steps. Every strategic decision without incentives thus depends only on the skill grid, the effort cost $\lambda/\gamma$, the AI accuracy $\PAI$, and $\kappa$.

\paragraph{Low-skill trap.}
For a worker who starts at level $p_0<a_N$, the closed form yields three regions of AI accuracy. The worker solves at $p_0$ if $\PAI\le\hat a(p_0)-\lambda/\gamma$. Everyone, including the expert, accepts if $\PAI>a_N-\lambda/\gamma$. In between,
\[
\hat a(p_0)-\lambda/\gamma<\PAI\le a_N-\lambda/\gamma,
\]
solving pays on its own at the top level, but the worker at $p_0$ accepts and never gets there. Because AI accuracy is a probability, the feasible trap region is
\[
\left(\hat a(p_0)-c,\ a_N-c\right]\cap[0,1],
\qquad c=\lambda/\gamma.
\]
Its width is
\[
L(c)=
\left[
\min\{1,a_N-c\}
-\max\{0,\hat a(p_0)-c\}
\right]^+.
\]
When the interval is not clipped by the probability boundaries, this reduces to $a_N-\hat a(p_0)$ and is independent of the effort cost and decay probability. This condition holds throughout the sampled design. For $p_0<a_N$, the unclipped width is positive at every finite horizon, shrinks as $\kappa$ grows, and tends to zero as the horizon becomes unbounded. Across the sampled grids, learning probabilities and horizons, its width at the lowest level ranges from $.003$ to $.60$. Outside this parameter range, clipping can make the feasible width depend on effort or eliminate the trap. Row~5 confirms that wherever the worker accepts at the two lowest levels, these levels have the same lifetime value.

\paragraph{Expert deskilling.}
Row~6 confirms that the expert accepts at every level and that lifetime value is the same at every level exactly when $a_N<\PAI+\lambda/\gamma$, for every horizon, learning probability, decay probability, and grid. The boundary $a_N-\lambda/\gamma$ is the vertical line in Figure~\ref{fig:who-solves}B in every configuration.

\paragraph{Effort incentive.}
Row~8 confirms that the effort incentive turns the myopic worker's choice into appropriate reliance. Row~9 extends Section~\ref{sec:effort-incentive} to strategic workers. Under the effort incentive, solving pays $\gamma p$ in expectation, so the closed form applies with $\lambda=0$: the strategic worker solves at $a_n$ if and only if $\hat a_n\ge\PAI$, and all levels accept if and only if $a_N<\PAI$. The effort incentive thus shifts both boundaries up by $\lambda/\gamma$ but leaves the unclipped width of the trap unchanged. A novice at $p_0$ with foresight and paid effort still accepts whenever $\hat a(p_0)<\PAI\le a_N$, although becoming an expert would pay. Foresight and the effort incentive together therefore do not close the low-skill trap either.

\paragraph{Learning and target incentives.}
Section~\ref{sec:dynamic-incentives} designs incentives for myopic workers and argues that they also move strategic workers (Appendix~\ref{app:incentives}). Rows~10 and~11 confirm this in every configuration. Under the learning incentive, the strategic worker solves at every level, and under a target incentive at every level up to the target. Row~13 shows that strategic workers never cost more in the long run than myopic workers under the same target. They solve at least as often, sometimes also just above the target, and can thereby reach levels at which payments end.

\paragraph{Acquiring a target.}
Row~12 checks the myopic target policy for every target $p^\star=a_j$, starting at $a_1$. For $j<N$, payments end if and only if $\Delta u(a_{j+1})\le0$, and their expected total equals Eq.~\eqref{eq:app-acquisition-cost}. For $j=N$, payments end if and only if $\Delta u(a_N)\le0$. Otherwise, the long-run expected payment per task is
\[
\Delta u(a_j)\,\frac{\phi_d}{\phi_u+\phi_d}\quad\text{for }j<N,
\qquad
\Delta u(a_N)\quad\text{for }j=N.
\]
For a myopic worker starting at $a_1$, whether payments end depends on the target's position relative to the myopic threshold $\PAI+\lambda/\gamma$. If this threshold lies above $a_1$ and at or below $a_N$, the highest grid level below it is the smallest sufficient support cutoff. Larger targets induce the same bonus schedule and have the same cost, because the additional bonuses are zero. If $a_1\ge\PAI+\lambda/\gamma$, no support is needed. If $a_N<\PAI+\lambda/\gamma$, no target ends payments under continued participation.

\begin{figure}[h]
\centering
\includegraphics[width=.55\linewidth]{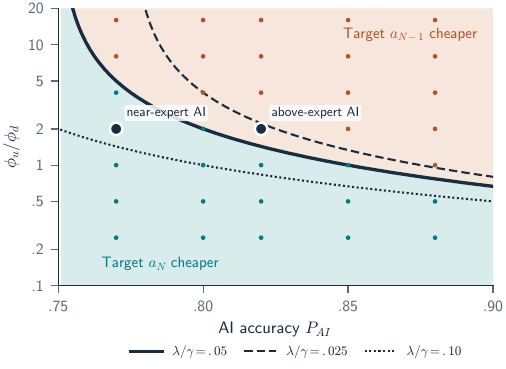}
\caption{Which retention target costs less in the long run, for a worker at $a_N=.80$ on the running-example grid. The higher target $a_N$ is cheaper below the boundary $\phi_u/\phi_d=\gamma s/\Delta u(a_N)$, the lower target $a_{N-1}$ above it. The solid boundary uses $\lambda/\gamma=.05$, the others $.025$ and $.10$. Dots are exact long-run payments for $\lambda/\gamma=.05$, colored by the cheaper target. The two marked points are the running example with the near-expert AI ($\PAI=.77$) and the above-expert AI ($\PAI=.82$), both with $\phi_u/\phi_d=2$.}
\label{fig:rob-retention}
\end{figure}

\paragraph{Retaining a target.}
Row~14 confirms the long-run payments of the targets $a_N$ and $a_{N-1}$ for a worker starting at $a_N$, and the condition $\Delta u(a_N)(\phi_u+\phi_d)<\Delta u(a_{N-1})\phi_d$ under which the higher target costs less. Unlike the other results, the comparison itself depends on the parameters. Among the 29{,}772 configurations with something to retain, $\Delta u(a_N)>0$, the higher target costs less in 9{,}907 ($33\%$). Figure~\ref{fig:rob-retention} shows the boundary for the running-example grid. The higher target is cheaper when $\PAI$ lies only slightly above the expert boundary $a_N-\lambda/\gamma$, so that $\Delta u(a_N)$ is small, and when skill decays quickly relative to how fast it is regained. For equally spaced levels with step $s$, the condition reads $\phi_u/\phi_d<\gamma s/\Delta u(a_N)$. With the above-expert AI ($\PAI=.82$), the running example lies on the other side of the boundary: $\gamma s/\Delta u(.80)=.10/.07=1.43<\phi_u/\phi_d=2$, so the lower target $.70$ costs less in the long run, at $.06$ instead of $.07$ per task, although it preserves less skill.

\subsection{Policy regions across parameters}\label{app:rob-regions}

\begin{figure}[h]
\centering
\includegraphics[width=\linewidth]{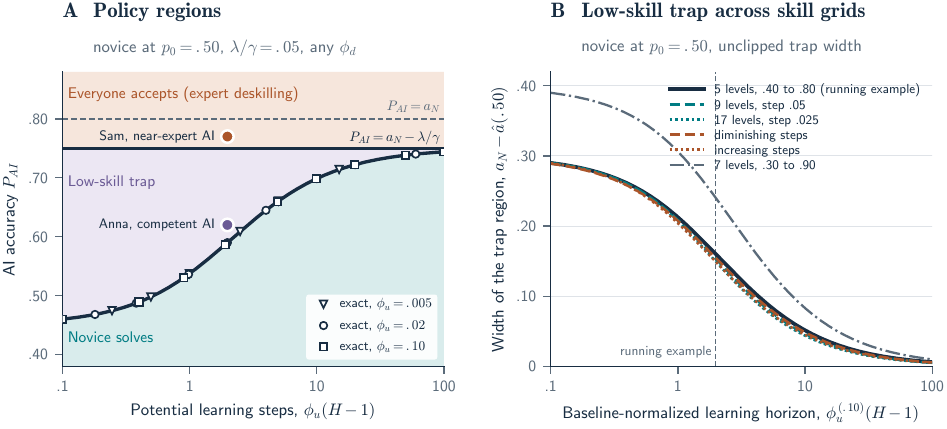}
\caption{The low-skill trap and expert deskilling across parameters. (A) Strategic choice of a novice at $p_0=.50$ on the running-example grid with $\lambda/\gamma=.05$, as a function of the expected number of learning steps over the horizon, $\kappa=\phi_u(H-1)$, and the AI accuracy $\PAI$. Below the curve, the novice solves. Between the curve and $\PAI=a_N-\lambda/\gamma$, solving would pay at the top level, but the novice accepts (low-skill trap). Above that line, everyone accepts (expert deskilling). The curve is the closed form of Eq.~\eqref{eq:app-strategic-rule}; markers are boundaries computed with the exact solver for three learning probabilities and several horizons, which fall on the same curve. The running example has $\kappa=.02\times99=1.98$. Decay $\phi_d$ does not affect the regions. (B) Width between the theoretical trap boundaries, $a_N-\hat a(.50)$, for six skill grids. The horizontal axis uses the baseline learning probability $\phi_u^{(.10)}$, with each grid's probability scaled as described in Appendix~\ref{app:rob-design}. The plotted width is measured before restricting the trap interval to feasible AI accuracies $\PAI\in[0,1]$. If both boundaries lie within $[0,1]$, this equals the feasible trap width and is independent of $\lambda/\gamma$. If a boundary lies outside $[0,1]$, only the intersection with $[0,1]$ is feasible; its width can be smaller and depend on $\lambda/\gamma$. Neither width depends on the decay probability $\phi_d$.}
\label{fig:rob-regions}
\end{figure}

Figure~\ref{fig:rob-regions}A summarizes the strategic choice of the novice for all learning probabilities, horizons, and decay probabilities at once. The running example lies in the middle of the range. With few expected learning steps, the boundary approaches the myopic one, $\PAI=.45$, and the trap covers almost every AI that would be worth beating. With many steps, the trap narrows toward the line $\PAI=a_N-\lambda/\gamma$, but it does not close. Anna is trapped with the competent AI because $\PAI=.62$ lies above the boundary at $\kappa=1.98$, which is $.59$. From about $\kappa=2.9$ on, for instance with $H=150$ instead of $100$, the boundary passes $.62$ and Anna would solve, but the trap would then apply to more accurate AIs. Sam deskills with the near-expert AI at every $\kappa$.

Figure~\ref{fig:rob-regions}B shows similar trap widths across the examined discretizations of $[.40,.80]$ after normalizing step probabilities by mean grid spacing. Finer grids and grids with diminishing or increasing steps give almost the same width as the running example. The width grows with the distance between the starting level and the top level, as for the wider grid on $[.30,.90]$, whose top level is higher. The low-skill trap is thus a matter of how far a novice is from the level at which independent work pays and of how many learning steps the remaining tasks allow, not of the particular grid.

\subsection{The running example under one-at-a-time variation}\label{app:rob-oat}

The qualitative results hold in every configuration, but whether a particular worker falls into one of the regions depends on the parameters. Table~\ref{tab:rob-oat} varies one parameter of the running example at a time and keeps $\PAI=.62$ for the competent and $.77$ for the near-expert AI.

\begin{table}[h]
\caption{Running-example outcomes when one parameter is varied at a time. \emph{Novice boundary}: largest $\PAI$ at which the strategic worker solves at $.50$. \emph{Expert boundary}: $a_N-\lambda/\gamma$. \emph{Trap width}: their difference. \emph{Anna trapped}: a strategic Anna at $.50$ accepts with the competent AI. \emph{Sam deskills}: a strategic Sam at $.80$ accepts with the near-expert AI. \emph{Bridging target}: cheapest target that ends payments for a novice starting at $a_1$ with the competent AI, and its expected total cost. \emph{Retention}: long-run payment per task to retain $a_N$ or $a_{N-1}$ with the near-expert AI; --- if Sam solves without an incentive. Strategic workers use $H=100$ unless stated otherwise. On the grid with 9 levels, $\phi_u=.04$ and $\phi_d=.02$. The grids with diminishing and increasing steps keep the running-example values $\phi_u=.02$ and $\phi_d=.01$.}
\label{tab:rob-oat}
\small
\begin{tabular}{@{}lccccccccc@{}}
\toprule
& \multicolumn{3}{c}{Boundaries in $\PAI$} & & & \multicolumn{2}{c}{Bridging target} & \multicolumn{2}{c}{Retention}\\
\cmidrule(lr){2-4}\cmidrule(lr){7-8}\cmidrule(l){9-10}
Variation & Novice & Expert & Trap width & Anna trapped & Sam deskills & Target & Cost & $a_N$ & $a_{N-1}$\\
\midrule
Running example & $.590$ & $.75$ & $.160$ & yes & yes & $.60$ & $25.50$ & $.020$ & $.040$\\
$\lambda/\gamma=.025$ & $.615$ & $.775$ & $.160$ & yes & no & $.60$ & $21.75$ & \multicolumn{2}{c}{---}\\
$\lambda/\gamma=.10$ & $.540$ & $.70$ & $.160$ & yes & yes & $.70$ & $34.00$ & $.070$ & $.057$\\
$\phi_u=.01$ & $.537$ & $.75$ & $.213$ & yes & yes & $.60$ & $51.00$ & $.020$ & $.060$\\
$\phi_u=.05$ & $.660$ & $.75$ & $.090$ & no & yes & $.60$ & $10.20$ & $.020$ & $.020$\\
$\phi_d=.005$ & $.590$ & $.75$ & $.160$ & yes & yes & $.60$ & $25.50$ & $.020$ & $.024$\\
$\phi_d=.02$ & $.590$ & $.75$ & $.160$ & yes & yes & $.60$ & $25.50$ & $.020$ & $.060$\\
$H=50$ & $.536$ & $.75$ & $.214$ & yes & yes & $.60$ & $25.50$ & $.020$ & $.040$\\
$H=200$ & $.645$ & $.75$ & $.105$ & no & yes & $.60$ & $25.50$ & $.020$ & $.040$\\
$H=1000$ & $.722$ & $.75$ & $.028$ & no & yes & $.60$ & $25.50$ & $.020$ & $.040$\\
9 levels, step $.05$ & $.597$ & $.75$ & $.153$ & yes & yes & $.65$ & $21.75$ & $.020$ & $.023$\\
diminishing steps & $.577$ & $.75$ & $.173$ & yes & yes & $.59$ & $26.00$ & $.020$ & $.027$\\
increasing steps & $.582$ & $.75$ & $.168$ & yes & yes & $.58$ & $38.00$ & $.020$ & $.047$\\
\bottomrule
\end{tabular}
\end{table}

\paragraph{Low-skill trap.}
Anna remains trapped under every variation except faster learning ($\phi_u=.05$) and longer horizons ($H\ge200$). In these cases the boundary rises above $.62$, because Anna expects enough learning steps for practice to pay. The trap width stays positive in every row, so the trap then applies to AIs between the new boundary and $.75$. Decay and the effort cost leave the width unchanged; the effort cost moves both boundaries by the same amount.

\paragraph{Expert deskilling.}
Sam deskills under every variation except the lower effort cost $\lambda/\gamma=.025$. Then $a_N-\lambda/\gamma=.775$ exceeds $.77$, Sam solves without an incentive, and nothing needs to be retained. No horizon, learning probability, decay probability, or grid on $[.40,.80]$ changes Sam's choice.

\paragraph{Incentive costs.}
The bridging target depends on the effort cost and the grid, because it is the highest level below the myopic threshold $.62+\lambda/\gamma$. With $\lambda/\gamma=.10$, a target of $.60$ no longer ends payments, and the organization must support practice up to $.70$, at a higher total cost. The total cost scales with $1/\phi_u$ and does not depend on $\phi_d$ or $H$. The higher retention target costs less in every row with something to retain, except that both targets cost the same with $\phi_u=.05$ and the lower target costs less with $\lambda/\gamma=.10$. In both cases, the condition of Appendix~\ref{app:incentives} does not hold strictly, because $\Delta u(a_N)$ is large relative to the gap one level down or because skill is regained quickly. The claim that a higher target can require less long-run spending holds in the running example and in most variations, but it is a possibility, not a general property, as Section~\ref{sec:targets} states.

\paragraph{What the analysis does not vary.}
The analysis varies parameters within the model, not its structure. Skill still moves at most one level per task, learning requires independent solving, the AI's accuracy is fixed, and workers know their own accuracy. Unequally spaced grids capture learning whose returns change with skill. Learning and decay probabilities that differ between levels leave the strategic decisions unchanged in form (Appendix~\ref{app:closed-form}), but the incentive costs then depend on the probability at each supported level. Section~\ref{sec:limitations} discusses these assumptions.

\end{document}